\documentclass[conference]{IEEEtran}
\IEEEoverridecommandlockouts
\usepackage{cite}
\usepackage{amsmath,amssymb,amsfonts}
\usepackage{graphicx}
\usepackage{textcomp}
\usepackage{xcolor}
\usepackage{booktabs}

\usepackage{mathtools}
\usepackage{ascmac}
\usepackage{url}
\usepackage{algorithm}
\usepackage{algpseudocode}
\usepackage{algorithmicx}

\usepackage{enumitem}

\def\BibTeX{{\rm B\kern-.05em{\sc i\kern-.025em b}\kern-.08em
    T\kern-.1667em\lower.7ex\hbox{E}\kern-.125emX}}
\begin{document}

\title{Emulation of Complex Matrix Multiplication based on the Chinese Remainder Theorem
}


\author{%
   \IEEEauthorblockN{%
       Yuki Uchino\IEEEauthorrefmark{1}, %
       Qianxiang Ma\IEEEauthorrefmark{2}, %
       Toshiyuki Imamura\IEEEauthorrefmark{3}, %
       Katsuhisa Ozaki\IEEEauthorrefmark{4}, and %
       Patrick Lars Gutsche\IEEEauthorrefmark{5}%
   }
   \IEEEauthorblockA{\IEEEauthorrefmark{1}%
       \textit{RIKEN Center for Computational Science}, Hyogo, Japan \\
       yuki.uchino.fe@riken.jp
   }
   \IEEEauthorblockA{\IEEEauthorrefmark{2}%
       \textit{RIKEN Center for Computational Science}, Hyogo, Japan \\
       qianxiang.ma@riken.jp
   }
   \IEEEauthorblockA{\IEEEauthorrefmark{3}%
       \textit{RIKEN Center for Computational Science}, Hyogo, Japan \\
       imamura.toshiyuki@riken.jp
   }
   \IEEEauthorblockA{\IEEEauthorrefmark{4}%
       \textit{Department of Mathematical Science} \\
       \textit{Shibaura Institute of Technology}, Saitama, Japan \\
       ozaki@sic.shibaura-it.ac.jp%
   }
   \IEEEauthorblockA{\IEEEauthorrefmark{5}%
       \textit{Computer Science Department} \\
       \textit{Ecole Normale Superieure de Lyon}, Lyon, France \\
       patrick.gutsche@ens-lyon.fr%
   }
}

\maketitle

\begin{abstract}
Modern computing architectures feature low-precision matrix multiplication units that achieve substantially higher throughput than their high-precision counterparts. 
Motivated by this architectural trend, the emulation of high-precision matrix multiplication using low-precision hardware has attracted significant interest in the high-performance computing community. 
Ozaki, Uchino, and Imamura proposed the Ozaki-II scheme as a general framework for emulating matrix multiplication. 
Building on this framework, Uchino, Ozaki, and Imamura developed high-performance and power-efficient techniques for emulating single- and double-precision real matrix multiplication on INT8 matrix engines. 
Extending this line of research, the present study proposes high-performance emulation methods for single- and double-precision complex matrix multiplication on INT8 matrix engines, based on the Ozaki-II scheme. 
On an NVIDIA B200 GPU, the proposed methods achieve 4.4--6.5$\times$ and 4.0--5.6$\times$ speedups over the native single- and double-precision complex matrix multiplication routines from cuBLAS, respectively, for sufficiently large problem sizes.
When lower accuracy than that of the standard routines is acceptable, the proposed methods can operate at even higher speed. 
Conversely, with only a modest increase in computation time, they can deliver higher accuracy than that of the standard routines. 
These properties suggest that the proposed approach has the potential to serve as a default algorithm across a wide range of applications.
\end{abstract}


\section{Introduction}
\label{sec:Introduction}

\subsection{Background}
\label{subsec:Background}
General-purpose computing architectures have increasingly adopted low-precision matrix multiplication units in graphics processing units (GPUs), tensor processing units (TPUs), and neural processing units (NPUs). 
Representative examples include Tensor Cores in NVIDIA GPUs~\cite{tensorcore}, Matrix Cores in AMD GPUs~\cite{matrixcore}, Matrix Multiply Units in Google TPUs~\cite{tpu}, and Multiply-Accumulate arrays in Intel NPUs~\cite{intelXecore}.
These units offer substantially higher arithmetic throughput than that of high-precision units and have become central components of modern accelerator design. 
Table~\ref{tab:specs} shows the theoretical throughput performance for AMD~\cite{MI300A,MI300X,MI325X,MI350X,MI355X} and NVIDIA GPUs~\cite{A100,H100,Blackwell,RTX5080}.
In recent years, low-precision matrix multiplication units have also begun to appear as standard features even in laptop-class processors~\cite{copilot+pc,SME}, further broadening their availability.
While these units were originally motivated by machine-learning workloads, their architectural characteristics also present an opportunity to accelerate high-precision computations by emulating such computations on low-precision hardware. 
Consequently, emulation-based high-precision matrix multiplication has emerged as an active research topic in the high-performance computing community~\cite{brower2025mixedprecisionabinitiotensor,abdelfattah2025analysisfloatingpointmatrixmultiplication,schwarz2025guaranteed,mukunoki2025sparseiterativesolversusing,dongarra2024hardwaretrendsimpactingfloatingpoint,11196192,11196413}.

\begin{table}[htbp]\centering
\caption{Performance (in TFLOPS or TOPS where appropriate) of AMD and NVIDIA GPUs~\cite{MI300A,MI300X,MI325X,MI350X,MI355X,A100,H100,Blackwell,RTX5080}}
\label{tab:specs}
\begin{tabular}{l@{ }c@{\quad }c@{\quad }c@{\quad }c@{\quad }c}
\toprule
GPU & FP64 & FP32 & FP16 & FP8 & INT8 \\
\midrule
Instinct MI300A CDNA3 & 122.6 & 122.6 & 980.6  & 1961.2 & 1961.2 \\
Instinct MI300X CDNA3 & 163.4 & 163.4 & 1307.4 & 2614.9 & 2614.9 \\
Instinct MI325X CDNA3 & 163.4 & 163.4 & 1307.4 & 2614.9 & 2614.9 \\
Instinct MI350X CDNA4 & 72.1  & 144.2 & 2306.9 & 4613.7 & 4614   \\
Instinct MI355X CDNA4 & 78.6  & 157.3 & 2516.6 & 5033.2 & 5033.2 \\
Ampere A100 SXM & 19.5  & 19.5  & 312    & 624    & 624    \\
Hopper H100 SXM & 66.9  & 66.9  & 989.4  & 1978.9 & 1978.9 \\
Blackwell B200 SXM       & 37    & 75    & 2250   & 4500   & 4500   \\
Blackwell Ultra B300 SXM & 1.2   & 75    & 2250   & 4500   & 150   \\
GeForce RTX 5080 Blackwell & 0.88 & 56.3 & 112.6 & 450.2 & 450.2 \\
\bottomrule
\end{tabular}
\end{table}

\subsection{Contributions}
\label{subsec:Contributions}
A prior line of work introduced the Ozaki-I~\cite{ozaki2012error,ozaki2013generalization,mukunoki2020,mukunoki2025dgemmfp64arithmetic,ootomo2024dgemm,uchino2025Performance} and Ozaki-II~\cite{ozaki-scheme2,uchino_ozaki2} schemes as frameworks for emulating high-precision matrix multiplication using low-precision matrix engines, demonstrating high performance and power efficiency on contemporary GPUs. 
The Ozaki-I scheme has recently been incorporated into cuBLAS (from version 13.0 Update 2). 
This implementation supports complex matrix multiplication. 
However, for real-valued matrix multiplication, previous studies have reported that the Ozaki-II scheme achieves higher performance than Ozaki-I. 
This suggests that a complex extension of the Ozaki-II scheme would likewise outperform the complex Ozaki-I approach. 
Despite this expectation, no complex-valued formulation of the Ozaki-II scheme has been developed. 
This absence is notable because complex matrix multiplication is a key computational kernel in many high-performance computing applications and its performance can dominate the overall runtime of large-scale scientific simulations.

To address this gap, the present work extends the Ozaki-II scheme to single- and double-precision complex matrix multiplication (CGEMM and ZGEMM, respectively). We develop emulation algorithms for CGEMM and ZGEMM by combining the Ozaki-II scheme with a Karatsuba-based formulation of complex arithmetic. 
The developed methods maintain numerical accuracy comparable to that of standard CGEMM and ZGEMM while effectively exploiting the high throughput of low-precision matrix engines. 
In addition, several improvements applicable to existing single- and double-precision general matrix multiplication (GEMM) emulation techniques are introduced, further enhancing the efficiency of the Ozaki schemes in the real domain.
The proposed algorithms are released as a portable library that supports both CUDA and HIP backends, enabling execution on NVIDIA and AMD GPUs. 
Performance is evaluated on an NVIDIA GH200 Grace Hopper Superchip, an NVIDIA B200 Blackwell GPU, an NVIDIA GeForce RTX 5080 Blackwell GPU, and an AMD Instinct MI300X GPU and compared against cuBLAS and hipBLAS implementations. 
For sufficiently large matrix sizes, the proposed methods surpass both vendor-optimized libraries and, in particular, achieve higher throughput than that of cuBLAS Ozaki-I-based ZGEMM emulation, demonstrating the performance potential of Ozaki-II-based emulation for CGEMM and ZGEMM.
We also provide performance models for the proposed methods. The predicted throughput closely aligns with the measurement results across all tested configurations.

The remainder of this paper is organized as follows. 
Section~\ref{sec:Ozaki-II Scheme for Real Matrix Multiplication} introduces the Ozaki-II scheme for real-valued matrix multiplication.
Section~\ref{sec:Proposed Methods} presents the proposed efficient methods for emulating CGEMM and ZGEMM.
Section~\ref{sec:Numerical Results} shows numerical results comparing the proposed emulation with vender-optimized GEMM routines.
Section~\ref{sec:Conclusion} presents the concluding remarks.
Note that in this paper, we do not consider matrix shapes for which performance becomes memory-bound, such as small matrices or very skinny matrices, because low-precision matrix engines do not achieve high throughput on such workloads.

\section{Ozaki-II Scheme for Real Matrix Multiplication}
\label{sec:Ozaki-II Scheme for Real Matrix Multiplication}
The Ozaki-II scheme is an emulation method for matrix multiplication based on a large-integer multiplication technique using the Chinese remainder theorem (CRT).
The scheme for the multiplication of matrices $A \in \mathbb{R}^{m \times k}$ and $B \in \mathbb{R}^{k \times n}$ consists of the following three steps:
\begin{enumerate}
\item\label{step:toint} Convert $A$ and $B$ to $A' \in \mathbb{Z}^{m \times k}$ and $B' \in \mathbb{Z}^{k \times n}$, respectively, by applying diagonal scaling and truncation.
\item\label{step:crt} Compute the integer matrix product $C' \approx A'B'$ based on the CRT.
\item Convert $C'$ back to floating-point format $C \in \mathbb{R}^{m \times n}$ via inverse scaling.
\end{enumerate}
The integer matrix multiplication in step~\ref{step:crt} is performed by decomposing the elements of $A'$ and $B'$ into residues with respect to a predefined set of pairwise coprime moduli.
Let $p_1, p_2, \dots, p_N \in \mathbb{N}$ denote these moduli and define $P := \prod_{\ell = 1}^N p_\ell$. 
Let $\mathrm{mod}(x,p_\ell)$ for $x \in \mathbb{Z}$ denote the symmetric remainder of $x/p_\ell$ (i.e., $\mathrm{mod}(x,p_\ell) = x - p_\ell \cdot \mathrm{round}(x/p_\ell)$, where $\mathrm{round}(\cdot)$ rounds its input to the nearest integer).
Each element of $A'$ and $B'$ is mapped to its residue representation modulo each $p_\ell$ as
\begin{equation}
(A'_\ell)_{ij} := \bmod(a'_{ij}, p_\ell),\quad
(B'_\ell)_{ij} := \bmod(b'_{ij}, p_\ell).
\end{equation}
Then, the corresponding matrix products are computed independently for each $p_\ell$ as
\begin{equation}
C'_\ell := A'_\ell B'_\ell.
\end{equation}
This procedure yields a set of partial results
\begin{equation}\label{eq:matmul}
C'_\ell \equiv A'B' \pmod{p_\ell}, \quad \ell = 1,\dots,N.
\end{equation}
The final integer matrix $C'$ is then reconstructed from $C'_\ell$ by applying the CRT as 
\[
C' := \sum_{\ell=1}^N \frac{P}{p_\ell} q_\ell\cdot C'_\ell.
\]
Then, we obtain
\[
C' \equiv A'B' \pmod{P},
\]
namely, $c'_{ij} = \mathrm{mod}\left((A'B')_{ij},P \right) + zP$ for $z \in \mathbb{Z}$.
If the condition
\begin{equation}\label{eq:cond}
    2\sum_{h = 1}^k|a'_{ih}||b'_{hj}| < P
\end{equation}
is satisfied for all $(i,j)$ pairs, the only possible value of $c'_{ij}$ is $\mathrm{mod}\left((A'B')_{ij}, P \right)$ and the CRT guarantees that the reconstructed $C'$ is unique.
Therefore, the scaling vector used in step~\ref{step:toint} must be chosen such that the bound~\eqref{eq:cond} is ensured for $A'$ and $B'$.
Increasing the number of moduli $p_\ell$ enlarges $P$, thereby relaxing the uniqueness condition.
However, this also increases the number of matrix multiplications required, leading to higher computational cost.

In previous work~\cite{uchino_ozaki2}, an algorithm was proposed in which each modular matrix multiplication~\eqref{eq:matmul} is executed using INT8 matrix engines.
Algorithm~\ref{alg:gemmul8} shows the outline.
Hereafter, we assume that $k  \le 2^{17}$ to perform INT8 matrix multiplication without error.
In~\cite{uchino_ozaki2}, step~V-\ref{alg:mod} of Algorithm~\ref{alg:gemmul8} uses the standard (non-symmetric) modulo operation that returns values in $[0,\, p_\ell)$ and $E'_\ell$ are UINT8 matrices.
In DGEMM emulation, step~V-\ref{alg:accumulation} of Algorithm~\ref{alg:gemmul8} evaluates $S = \sum_{\ell=1}^{N} P/p_\ell\cdot q_\ell\cdot E'_\ell$ using the unevaluated sum $S_1 + S_2$ as
\begin{equation}\label{eq:CRTaccumulation}
S \approx S_1 + S_2,\quad S_1 := \sum_{\ell=1}^N s_{\ell 1} E'_\ell,\quad S_2 := \sum_{\ell=1}^N s_{\ell 2} E'_\ell,
\end{equation}
where $s_{\ell 1}$ contains the upper $(53 - 8 - \lceil \log_2 N \rceil)$ bits of $P/p_\ell\cdot q_\ell$, $s_{\ell 2}$ stores the remaining lower bits rounded to double precision, and $s_{\ell 1} + s_{\ell 2} \approx P/p_\ell\cdot q_\ell$.
In~\eqref{eq:CRTaccumulation}, the value of $s_{\ell 1}$ is limited by the precision of the double-precision significand.
Since each element of $E'_\ell$ is an 8-bit unsigned integer, the product $s_{\ell 1} E'_\ell$ requires 8 bits of significand to represent the contribution of $E'_\ell$.
Moreover, accumulating $N$ such terms requires an additional $\lceil \log_2 N \rceil$ bits to avoid overlap between partial sums.
Thus, the high-order part $s_{\ell 1}$ may occupy at most $(53 - 8 - \lceil \log_2 N \rceil)$ bits of precision.
This bit allocation ensures that all products $s_{\ell 1} E'_\ell$ can be accumulated in $S_1$ without rounding error.
In contrast, the present work adopts a symmetric modulo operation in step~V-\ref{alg:mod} of Algorithm~\ref{alg:gemmul8}, ensuring that each $E'_\ell$ becomes INT8 from UINT8.
Consequently, the upper part $s_{\ell 1}$ of $P/p_\ell\cdot q_\ell$ can use $(53 - 7 - \lceil \log_2 N \rceil)$ bits instead of $(53 - 8 - \lceil \log_2 N \rceil)$ bits, gaining one additional effective bit compared to the previous method.
This slight but meaningful increase in representable precision improves the numerical robustness of the reconstruction step without changing the overall algorithmic structure.

\begin{algorithm}
\caption{Outline of Ozaki-II scheme using INT8 matrix engines for real matrix multiplication~\cite{uchino_ozaki2}. The function $\mathrm{trunc}(\cdot)$ returns the integer part of its input.\label{alg:gemmul8}}
\begin{algorithmic}
\Require $A \in \mathbb{R}^{m \times k}$, $B \in \mathbb{R}^{k \times n}$, $N \in \mathbb{N}$
\Ensure $C \approx AB \in \mathbb{R}^{m \times n}$
\end{algorithmic}
\begin{description}
\item[Step I] Determine the moduli $p_\ell \le 256$ for $\ell = 1, \dots, N$.
\item[Step II] Determine modular multiplicative inverses $q_\ell$ of $P/p_\ell$ modulo $p_\ell$ for $\ell = 1, \dots, N$.
\item[Step III] Determine the scaling vectors $\mu \in \mathbb{R}^m$ and $\nu \in \mathbb{R}^n$.
\item[Step IV] Convert $A$ and $B$ to $A' \in \mathbb{Z}^{m \times k}$ and $B' \in \mathbb{Z}^{k \times n}$:
\begin{enumerate}[label=\roman*),ref=\roman*]
    \item $a'_{ij} := \mathrm{trunc}(a_{ij} \cdot \mu_i)$
    \item $b'_{ij} := \mathrm{trunc}(b_{ij} \cdot \nu_j)$
\end{enumerate}
\item[Step V] Compute $C' \approx A'B'$ based on the CRT:
\begin{enumerate}[label=\roman*),ref=\roman*]
    \item $(A'_\ell)_{ij} := \bmod(a'_{ij}, p_\ell)$ for $\ell = 1, \dots, N$
    \item $(B'_\ell)_{ij} := \bmod(b'_{ij}, p_\ell)$ for $\ell = 1, \dots, N$
    \item $D'_\ell := A'_\ell B'_\ell$ using INT8-GEMM for $\ell = 1, \dots, N$
    \item\label{alg:mod} $(E'_\ell)_{ij} := \bmod(d'_{ij}, p_\ell)$ for $\ell = 1, \dots, N$
    \item\label{alg:accumulation} $S := \sum_{\ell=1}^N P/p_\ell \cdot q_\ell \cdot E'_\ell$
    \item $c'_{ij} := \bmod(s_{ij}, P)$
\end{enumerate}
\item[Step VI] Convert $C'$ back to floating-point format $C \in \mathbb{R}^{m \times n}$:
\begin{enumerate}[label=\roman*),ref=\roman*]
    \item $C := \mathrm{diag}(\mu)^{-1} C' \mathrm{diag}(\nu)^{-1}$
\end{enumerate}
\end{description}
\end{algorithm}

To satisfy the condition~\eqref{eq:cond} efficiently, the scaling vectors used in step~IV of Algorithm~\ref{alg:gemmul8} must be determined appropriately.
Previous work introduced two strategies for computing this scaling: a fast mode and an accurate mode.
The fast mode estimates the upper bound of $\sum_{h = 1}^k|a'_{ih}||b'_{hj}|$ via the Cauchy--Schwarz inequality and determines the scaling vector analytically based on this estimate.
This approach incurs minimal preprocessing cost, but the obtained bound tends to be overly conservative.
As a consequence, a larger $P$ (i.e., more moduli) is often required to satisfy the uniqueness condition, increasing the computational workload in step~V Algorithm~\ref{alg:gemmul8}.
In contrast, the accurate mode estimates the upper bound by performing an auxiliary matrix multiplication using INT8 matrix engines.
Although this mode requires additional computation during step~III of Algorithm~\ref{alg:gemmul8}, the direct estimation yields a much tighter bound than that obtained via the Cauchy--Schwarz inequality.
This typically enables the use of fewer moduli, reducing the number of modular matrix multiplications and potentially improving overall performance.

\section{Proposed Methods}
\label{sec:Proposed Methods}
This section presents the proposed extension of the Ozaki-II scheme to complex matrix multiplication, targeting high-performance execution on INT8 matrix engines.
Let $A = A_R + iA_I \in \mathbb{C}^{m \times k}$ for $A_R,A_I \in \mathbb{R}^{m \times k}$ and $B = B_R + iB_I \in \mathbb{C}^{k \times n}$ for $B_R,B_I \in \mathbb{R}^{k \times n}$. We compute $C = C_R + iC_I := AB$, where $C_R,C_I \in \mathbb{R}^{m \times n}$.
A standard approach for handling complex matrix multiplication is to employ the following representation:
\begin{equation}\label{eq:complex-matrix-representation}
\hat{M} :=
\begin{pmatrix}
    M_R & -M_I\\
    M_I & M_R
\end{pmatrix}
\end{equation}
for $M \in \{A,B,C\}$.
This representation forms the basis for both steps~\ref{step:toint} and~\ref{step:crt} in the complex Ozaki-II scheme.

\subsection{INT8 Complex Matrix Multiplication}
\label{subsec:INT8 Complex Matrix Multiplication}
\begin{figure}[htbp]\centering
\includegraphics[width=\hsize]{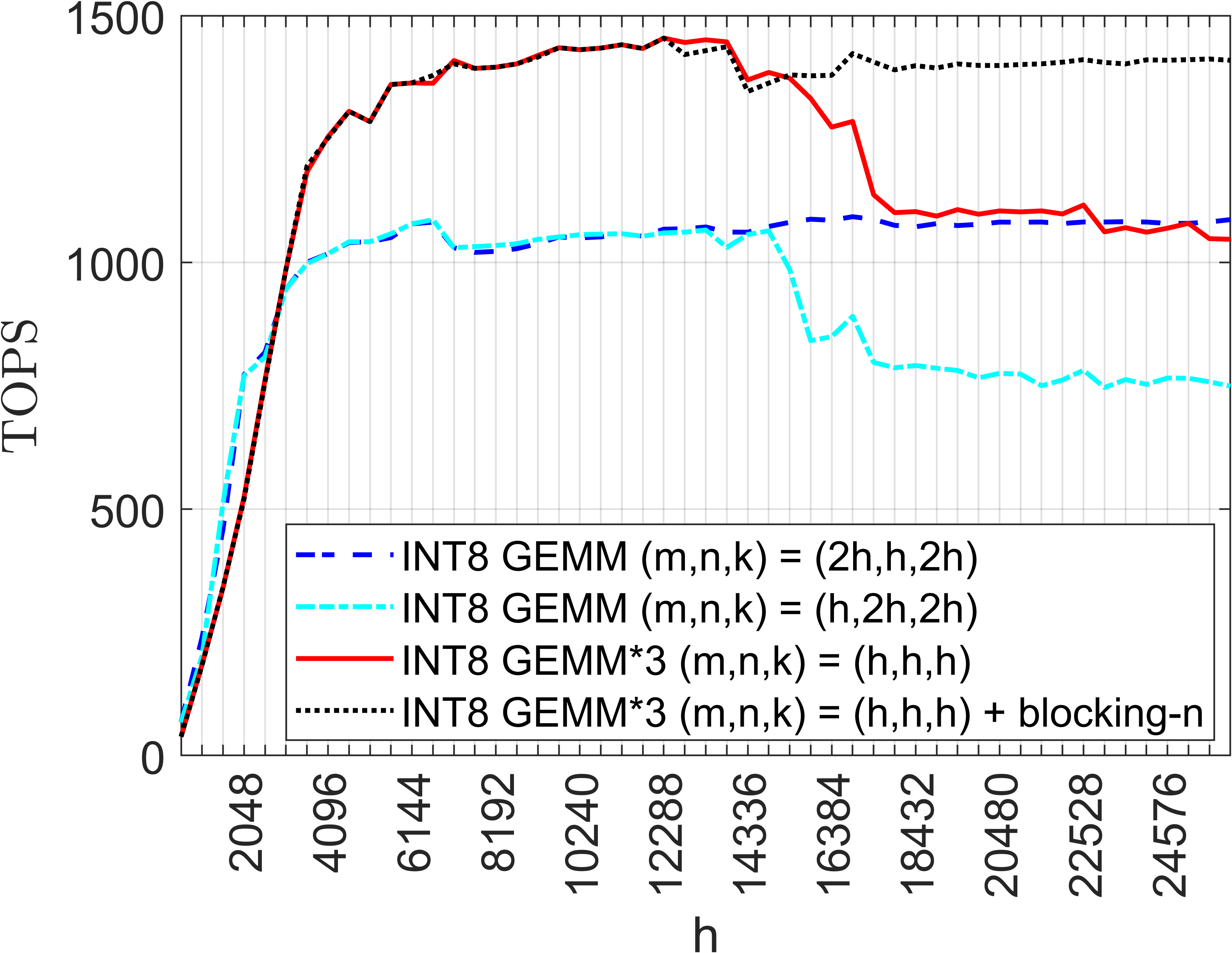}
\caption{Performance comparison of four INT8-based GEMM strategies on NVIDIA H100 NVL (CUDA Toolkit~12.8.61).
The first two methods use a single INT8 matrix multiplication with dimensions $(2h, h, 2h)$ and $(h, 2h, 2h)$, respectively.
The third method performs three INT8 multiplications of size $(h, h, h)$. The fourth method applies the same three-multiplication scheme but introduces blocking along the $n$ dimension.}
\label{fig:int8-complex-gemm-perf}
\end{figure}

Using~\eqref{eq:complex-matrix-representation}, $C_R$ and $C_I$ are obtained using
\begin{equation}\label{eq:complex-gemm-1}
\begin{pmatrix}
    C_R\\
    C_I
\end{pmatrix}
=
\begin{pmatrix}
    A_R & -A_I\\
    A_I & A_R
\end{pmatrix}
\begin{pmatrix}
    B_R\\
    B_I
\end{pmatrix}
\end{equation}
or
\begin{equation}\label{eq:complex-gemm-2}
\begin{pmatrix}
    C_I & C_R
\end{pmatrix}
=
\begin{pmatrix}
    A_I & A_R
\end{pmatrix}
\begin{pmatrix}
    B_R & -B_I\\
    B_I & B_R
\end{pmatrix}.
\end{equation}
The formulations in \eqref{eq:complex-gemm-1} and \eqref{eq:complex-gemm-2} allow complex matrix multiplication to be computed using a single real GEMM applied to an expanded matrix.
For the formulations, the applicability condition on the inner dimension changes from $k \le 2^{17}$ to $k \le 2^{16}$.
Another standard approach for handling complex matrix multiplication is the Karatsuba algorithm.
The Karatsuba algorithm expresses the complex product using three real matrix multiplications as 
\begin{equation}
    C_R := D-E,\quad
    C_I := F-D-E
\end{equation}
for
\begin{equation}\label{eq:karatsuba}
    \begin{cases}
        D := A_R B_R,\\
        E := A_I B_I,\\
        F := (A_R + A_I)(B_R + B_I),
    \end{cases}
\end{equation}
without forming enlarged matrices.
We evaluated the three approaches based on~\eqref{eq:complex-gemm-1}, \eqref{eq:complex-gemm-2}, and \eqref{eq:karatsuba}, respectively, on INT8 matrix engines.
Note that both $A_R + A_I$ and $B_R + B_I$ in~\eqref{eq:karatsuba} are reduced to the INT8 range by applying the modulo operation before the products are formed.
The performance results are summarized in Fig.~\ref{fig:int8-complex-gemm-perf}.
In addition to these three methods, we also tested a blocked variant of~\eqref{eq:karatsuba}.
This variant applies the Karatsuba formulation but partitions the output dimension $n$ into blocks of size 8192, thereby limiting the working set of each INT8 GEMM and preventing performance degradation for very large $n$.
For problem sizes up to $h \approx 2000$, the methods based on~\eqref{eq:complex-gemm-1} and~\eqref{eq:complex-gemm-2} are slightly faster.
When $h > 3000$, both the original~\eqref{eq:karatsuba} and the blocking variant of~\eqref{eq:karatsuba} achieve 1.2--1.4$\times$ higher throughput.
However, for $h > 15000$, the performance of the unblocked~\eqref{eq:karatsuba} deteriorates, while the blocked version maintains stable throughput.
Similar results were observed on other GPUs, such as the GH200 and B200.
Based on these observations, the proposed complex Ozaki-II scheme adopts the Karatsuba method with $n$-blocking for all modular INT8 complex matrix multiplications.

\subsection{Scaling Vector Determination}
\label{subsec:Scaling Vector Determination}
First, we convert $\hat{A}$ and $\hat{B}$ to integer matrices $\hat{A}'$ and $\hat{B}'$, respectively, as 
\[
\hat{a}'_{ij} := \mathrm{trunc}(\hat{a}_{ij} \cdot \mu_i),\quad
\hat{b}'_{ij} := \mathrm{trunc}(\hat{b}_{ij} \cdot \nu_j).
\]
We consider two strategies: a fast mode and an accurate mode.
From $|\hat{a}'_{ij}| \le |\hat{a}_{ij}| \cdot \mu_i$ and $|\hat{b}'_{ij}| \le |\hat{b}_{ij}| \cdot \nu_j$, 
\[
    \sum_{h=1}^k |\hat{a}'_{ih}||\hat{b}'_{hj}| \le \mu_i\nu_j \cdot \sum_{h=1}^k |\hat{a}_{ih}| |\hat{b}_{hj}|.
\]
In the fast mode, using the Cauchy--Schwarz inequality yields
\[
\sum_{h=1}^k |\hat{a}_{ih}| |\hat{b}_{hj}|
\le \|\hat{a}_{i,:}\|_2 \|\hat{b}_{:,j}\|_2.
\]
From~\eqref{eq:complex-matrix-representation}, the row norms of $\hat{A}$ satisfy $\|\hat{a}_{i,:}\|_2 = \|\hat{a}_{i+m,:}\|_2$ for $1 \le i \le m$, and the column norms of $\hat{B}$ satisfy $\|\hat{b}_{:,j}\|_2 = \|\hat{b}_{:,j+n}\|_2$ for $1 \le j \le n$.
Thus, the scaling values for rows $i$ and $i+m$ of $\hat{A}$ are identical; the same holds for columns $j$ and $j+n$ of $\hat{B}$.
Consequently, the scaling vectors reduce to $\mu \in \mathbb{R}^m$ and $\nu \in \mathbb{R}^n$.
The elements of $\mu$ and $\nu$ can then be determined in the same manner as in the real-valued Ozaki-II scheme~\cite{uchino_ozaki2}; that is, 
\begin{align*}
\mu_i &\le 2^{\lfloor \frac{\log_2 (P-1) - 1}{2} - 1 -\max\left(1,\ 0.5\cdot \log_2 \sum_h \hat{a}_{ih}^2 \right) \rfloor - \lfloor \log_2 \max_{h}|\hat{a}_{ih}| \rfloor},\\
\nu_j &\le 2^{\lfloor \frac{\log_2 (P-1) - 1}{2} - 1 -\max\left(1,\ 0.5\cdot \log_2 \sum_h \hat{a}_{ih}^2 \right) \rfloor - \lfloor \log_2 \max_{h}|\hat{b}_{hj}| \rfloor}.
\end{align*}
For any real number $x \in \mathbb{R}$, $\lfloor \log_2 |x| \rfloor$ represents the unit in the first place of $x$, and thus it can be obtained directly using the $\mathtt{ilogb}(\cdot)$ function.
Following~\cite{uchino_ozaki2}, the quantity $(\log_2 (P-1) - 1)/2 - 1$ is precomputed as $\mathcal{P}'_{\mathit{fast}} := \mathrm{single}(\log_2 (P-1)/2 - 1.5)$.
Furthermore, $\log_2 \|\hat{a}_{i,:}\|_2$ and $\log_2 \|\hat{b}_{:,j}\|_2$ can be upper-bounded by
$0.51 \cdot \mathtt{\_\_log2f}(\sum_{h=1}^{2k}\hat{a}_{ih}^2)$ and
$0.51 \cdot \mathtt{\_\_log2f}(\sum_{h=1}^{2k}\hat{b}_{hj}^2)$, respectively, using floating-point arithmetic in round-up mode.
Here, $\mathtt{\_\_log2f}$ denotes the fast built-in base-2 logarithm function provided by the CUDA Math API and the HIP math API.
According to NVIDIA's official CUDA documentation, the maximum absolute error of $\mathtt{\_\_log2f}(x)$ is $2^{-22}$ for $x \in [0.5, 2]$, and at most 2 units in the last place (ULPs) otherwise~\cite{cuda}.
The definition of ULP used in the CUDA documentation follows Muller~\cite{muller:inria-00070503}.
In addition, the AMD HIP documentation states that the maximum absolute error of $\mathtt{\_\_log2f}(x)$ is 1 ULP for $x \in [10^{-6}, 10^{6}]$~\cite{hip}, although the precise definition of ULP is not specified.
Based on these specifications, we assume that the bound $\mathtt{\_\_log2f}(x) \le (\log_2 x)/(1 - 4u)$, where $u$ is the unit roundoff of FP32 arithmetic.
If $x$ is not sufficiently close to $1$, it is first scaled to $x' := x / 2^{\lfloor \log_2 |x| \rfloor}$, after which the logarithm is evaluated as $\mathtt{\_\_log2f}(x') + \lfloor \log_2 |x| \rfloor$.
Based on the above considerations, for $\delta := \mathrm{single}(0.5/(1-4u))$ computed in round-down mode, the scaling vectors $\mu$ and $\nu$ are determined as follows:
\begin{align}
\mu_i &:= 2^{\lfloor \mathcal{P}'_{\mathit{fast}}-\max(1,\ \delta \cdot \mathtt{\_\_log2f}(\sum_h \hat{a}_{ih}^2) ) \rfloor - \lfloor \log_2 \max_{h}|\hat{a}_{ih}| \rfloor},\label{eq:mu-fast}\\
\nu_j &:= 2^{\lfloor \mathcal{P}'_{\mathit{fast}}-\max(1,\ \delta \cdot \mathtt{\_\_log2f}(\sum_h \hat{b}_{hj}^2) ) \rfloor - \lfloor \log_2 \max_{h}|\hat{b}_{hj}| \rfloor},\label{eq:nu-fast}
\end{align}
where $\mathcal{P}'_{\mathit{fast}} - \delta \cdot x$ for $x \in \mathbb{R}$ is computed using the fused multiply-add operation in round-down mode.

In the accurate mode, we first apply diagonal scaling to $A_R$, $A_I$, $B_R$, and $B_I$ so that the integer parts of the maximum absolute values in each row of $A_R$ and $A_I$ and in each column of $B_R$ and $B_I$ fit within 6 bits.
Specifically, we scale them as 
$\mathrm{diag}(\bar{\mu}) A_R$, 
$\mathrm{diag}(\bar{\mu}) A_I$, 
$B_R \mathrm{diag}(\bar{\nu})$, and 
$B_I \mathrm{diag}(\bar{\nu})$, 
and then round them up using the ceiling operation to obtain 7-bit or smaller integer matrices $\bar{A}_R$, $\bar{A}_I$, $\bar{B}_R$, and $\bar{B}_I$, respectively.
These matrices satisfy
\begin{alignat*}{2}
|\bar{\mu}_i (A_R)_{ij}| &\le (\bar{A}_R)_{ij},&\quad
|\bar{\mu}_i (A_I)_{ij}| &\le (\bar{A}_I)_{ij},\\
|(B_R)_{ij} \bar{\nu}_j| &\le (\bar{B}_R)_{ij},&\quad
|(B_I)_{ij} \bar{\nu}_j| &\le (\bar{B}_I)_{ij}.
\end{alignat*}
After this preprocessing, we compute the upper bound $\bar{C}_R + i\bar{C}_I$ of the INT8 matrix product $(\bar{A}_R + i\bar{A}_I)(\bar{B}_R + i\bar{B}_I)$ as follows:
\begin{align*}
    \bar{C}_I &:= \bar{A}_I\bar{B}_R + \bar{A}_R\bar{B}_I,\\
    \bar{C}_R &:= \bar{C}_I + (\bar{A}_R - \bar{A}_I)(\bar{B}_R - \bar{B}_I) = \bar{A}_R\bar{B}_R + \bar{A}_I\bar{B}_I,
\end{align*}
where no computing error occurs in $\bar{A}_R - \bar{A}_I$ and $\bar{B}_R - \bar{B}_I$ in INT8.
Then, the elements of $\mu$ and $\nu$ can be determined in the same manner as in the real-valued Ozaki-II scheme~\cite{uchino_ozaki2}; that is, 
\begin{align}
\mu_i &:= \bar{\mu}_i \cdot 2^{\lfloor \mathcal{P}'_{\mathit{accu}} - \delta \cdot \log_2(\max_{h} \max( (\bar{C}_R)_{ih}, (\bar{C}_I)_{ih} )) \rfloor},\label{eq:mu-accu}\\
\nu_j &:= \bar{\nu}_j \cdot 2^{\lfloor \mathcal{P}'_{\mathit{accu}} - \delta \cdot \log_2(\max_{h} \max( (\bar{C}_R)_{hj}, (\bar{C}_I)_{hj} )) \rfloor},\label{eq:nu-accu}
\end{align}
where $\mathcal{P}'_{\mathit{accu}} := \mathrm{single}(\log_2 (P-1)/2-0.5)$ and $\mathcal{P}'_{\mathit{accu}} - \delta \cdot x$ for $x \in \mathbb{R}$ is computed using the fused multiply-add operation in round-down mode.

\subsection{Performance Analysis}
\label{subsec:Performance Analysis}
To analyze the performance characteristics of the proposed methods, we employ a simple yet expressive performance model that captures both computation and memory transfer costs. Let $b$ denote the peak sustained memory bandwidth (in B/s) and $p$ denote the peak throughput of the INT8 matrix-multiplication engine (in operations per second, OPS). 
For a given operation, the memory-bound execution time is estimated as the required data movement divided by $b$ and the compute-bound time is obtained as the arithmetic workload divided by $p$.
Because even memory-bound kernels incur a non-negligible amount of arithmetic overhead, we include a correction term $c$.

\subsubsection{Fast Mode}
\label{subsubsec:Fast Mode}
Let the input matrices $A \in \mathbb{C}^{m \times k}$ and $B \in \mathbb{C}^{k \times n}$ be double-precision.
In step~1, both $A$ and $B$ are loaded twice:
once for computing the row-wise maximum absolute values of $\hat{A}$ and the column-wise maximum absolute values of $\hat{B}$, and once for computing the residues of $\hat{A}$ and $\hat{B}$.
Afterward, the scaling vectors and the INT8 matrices are stored.
Each scaling factor is a power of two, so only its exponent needs to be stored as an INT16.
Including the correction term $c$, which accounts for arithmetic overhead even in memory-bound operations,
the total execution time for step~1 can be estimated as $((32 + 3N + c)k + 2)(m+n)/b$.
In step~2, the INT8 GEMM kernels are executed $3N$ times.
The resulting $3N$ INT32 matrices are then loaded and converted into $N$ INT8 complex matrices.
Accounting for computational overhead $c$, the execution time for step~2 is estimated as $6Nmnk/p + (14N + c)mn/b$.
In step~3, the $N$ INT8 complex matrices are loaded and the CRT accumulation and final reduction are performed.
The scaling vectors are then loaded, the inverse scaling is applied, and finally the resulting double-precision complex matrix is stored.
With the correction term $c$, the execution time for step~3 is estimated as $(2(m+n) + (2N + 16 + c)mn)/b$.

Combining the costs derived for steps~1, 2, and 3, the total execution time for the proposed methods can be estimated as the sum of its memory-bound and compute-bound components.
Aggregating all terms yields the following expression for the overall execution time:
\[
\begin{multlined}
\frac{((3N+32+c)k + 4)(m+n) + (16N+16+2c)mn}{b}\\
+ \frac{6Nmnk}{p}.
\end{multlined}
\]
Similarly, when both $A$ and $B$ are single-precision complex matrices, the total execution time can be estimated as
\[
\begin{multlined}
\frac{((3N+16+c)k + 4)(m+n) + (16N+8+2c)mn}{b}\\
+ \frac{6Nmnk}{p}.
\end{multlined}
\]

\subsubsection{Accurate Mode}
\label{subsubsec:Accurate Mode}
In the accurate mode, an additional INT8 complex matrix multiplication is performed in step~1, which also increases the memory traffic due to extra stores of INT8 matrices and loads of INT32 matrices, as well as additional loads and stores of the scaling vectors.
The total execution time in this mode can therefore be estimated as
\[
\begin{multlined}
\frac{((35+3N + c)k+8)(m+n) + (16N + 40 + 2c)mn}{b}\\
 + \frac{6(N+1)mnk}{p},
\end{multlined}
\]
and for the single-precision case as
\[
\begin{multlined}
\frac{((19+3N + c)k+8)(m+n) + (16N + 32 + 2c)mn}{b}\\
 + \frac{6(N+1)mnk}{p}.
\end{multlined}
\]

\begin{figure}[htbp]\centering
\noindent
\begin{minipage}[b]{0.49\hsize}
\includegraphics[width=\hsize]{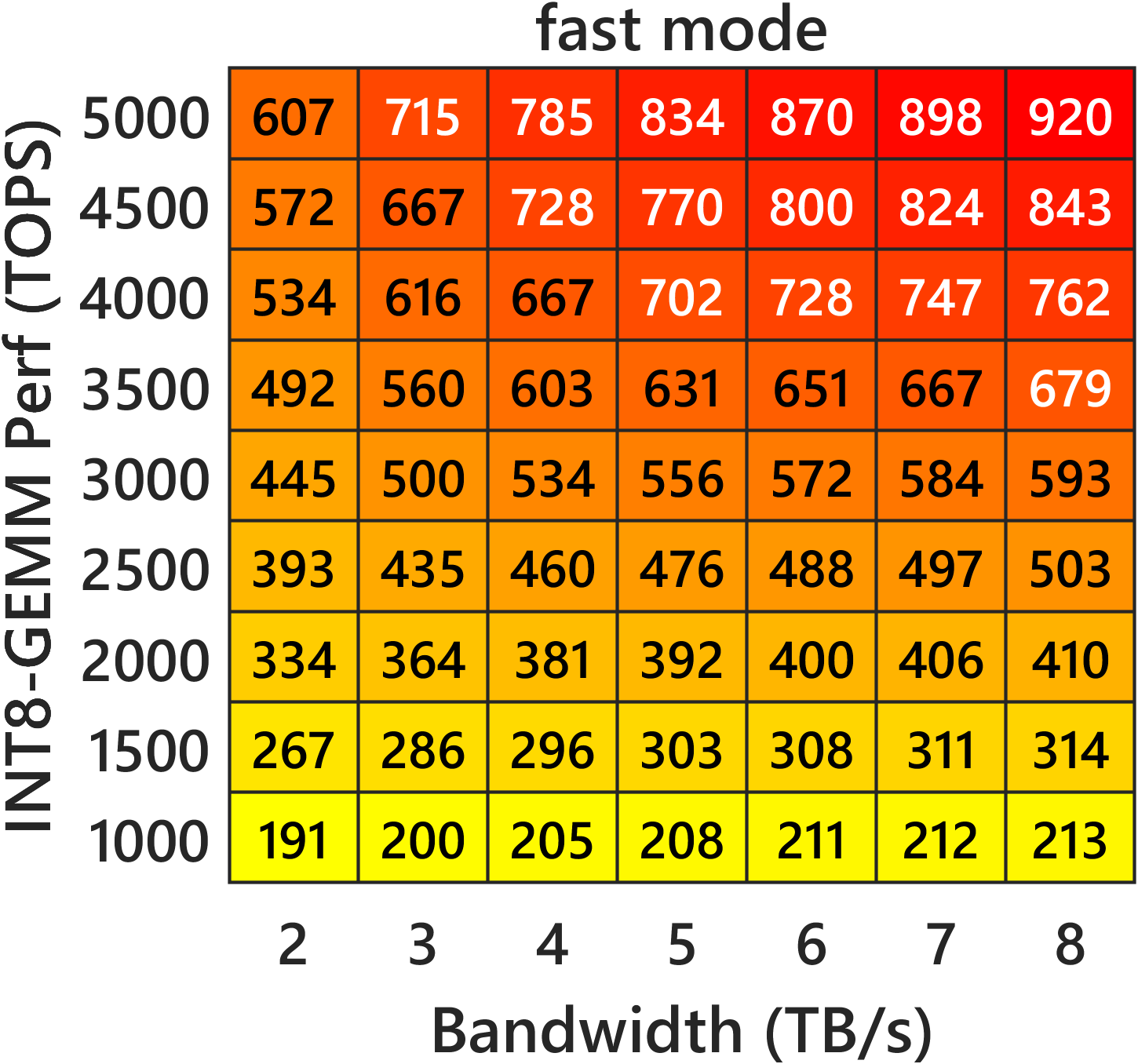}
\end{minipage}
\begin{minipage}[b]{0.49\hsize}
\includegraphics[width=\hsize]{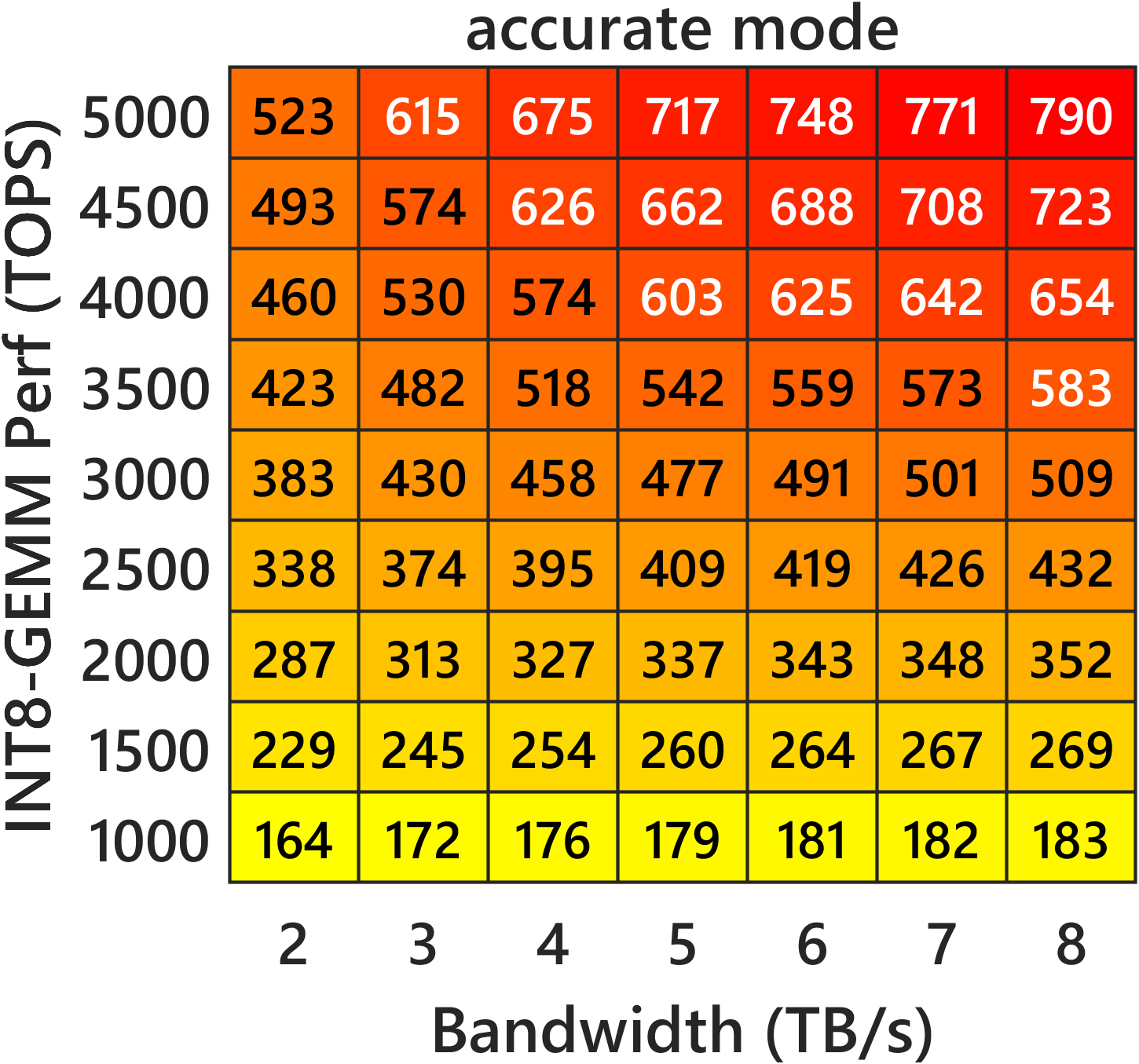}
\end{minipage}
\caption{Performance model heatmaps for single-precision complex matrix multiplication emulation.
The left panel corresponds to the fast mode and the right panel corresponds to the accurate mode.
The horizontal axis denotes the achievable memory bandwidth and the vertical axis denotes the achievable INT8 GEMM throughput.
The problem size is fixed at $m=n=k=16384$ and the correction term is set to $c=6$, equal to the number of moduli used in the emulation.
The color scale indicates the predicted throughput (in TFLOPS) of the proposed emulation.}
\label{fig:C_heatmap_perf-est-Ozaki2}
\end{figure}

\begin{figure}[htbp]\centering
\noindent
\begin{minipage}[b]{0.49\hsize}
\includegraphics[width=\hsize]{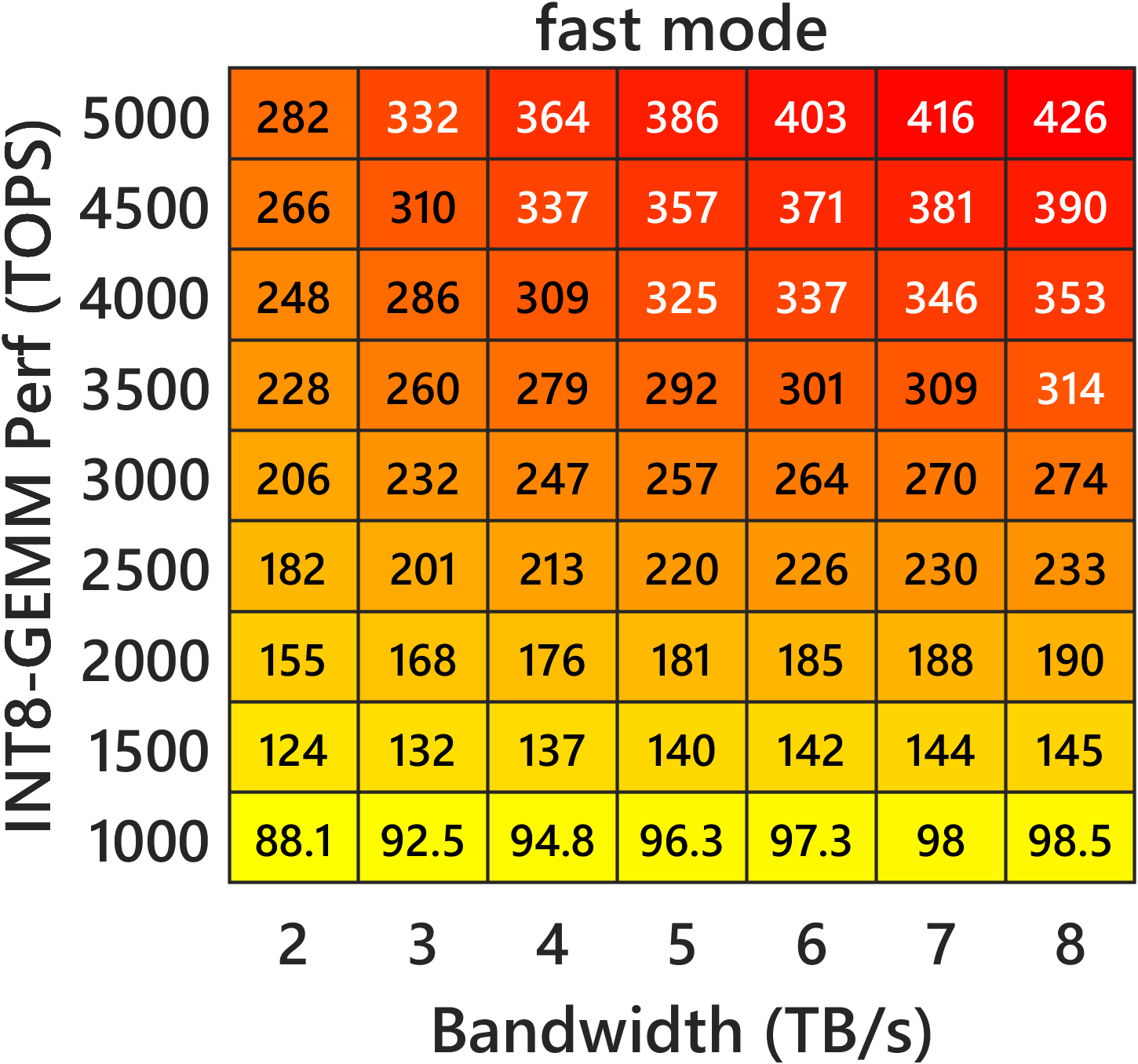}
\end{minipage}
\begin{minipage}[b]{0.49\hsize}
\includegraphics[width=\hsize]{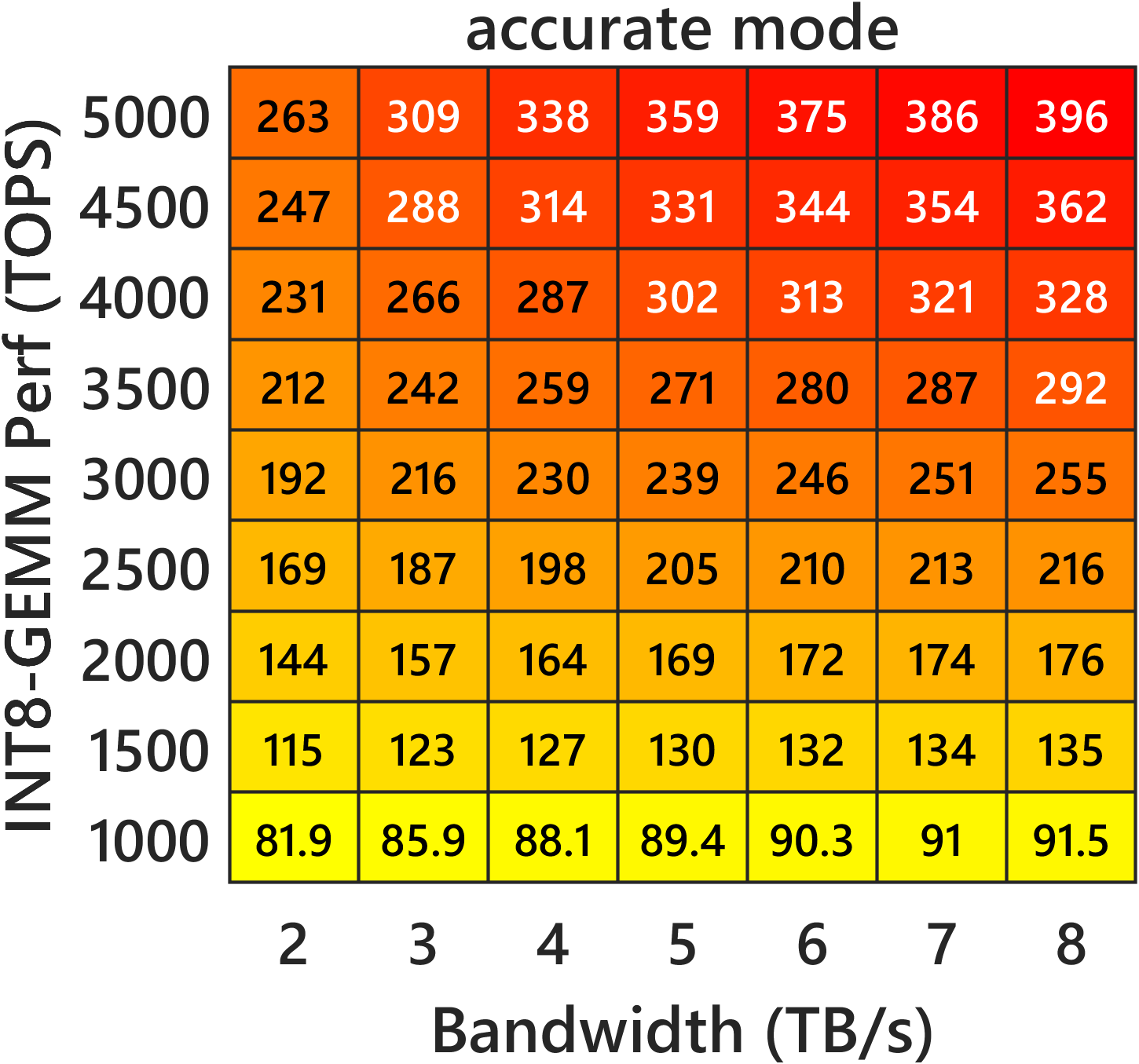}
\end{minipage}
\caption{Performance model heatmaps for double-precision complex matrix multiplication emulation.
The left panel corresponds to the fast mode and the right panel corresponds to the accurate mode.
The horizontal axis denotes the achievable memory bandwidth and the vertical axis denotes the achievable INT8 GEMM throughput.
The problem size is fixed at $m=n=k=16384$ and the correction term is set to $c=13$, equal to the number of moduli used in the emulation.
The color scale indicates the predicted throughput (in TFLOPS) of the proposed emulation.}
\label{fig:Z_heatmap_perf-est-Ozaki2}
\end{figure}

Figs.~\ref{fig:C_heatmap_perf-est-Ozaki2} and \ref{fig:Z_heatmap_perf-est-Ozaki2} show performance model heatmaps for the fast and accurate modes in single and double precision for $6$ and $13$ moduli, respectively.
The TFLOPS was computed as
\[
\frac{8mnk}{t} \cdot 10^{-12},
\]
where $t$ is the execution time in seconds.
For example, assuming an effective memory bandwidth of 2--4 TB/s and an observed INT8 GEMM kernel throughput of 1500 TFLOPS on the GH200, the model predicts that ZGEMM emulation in the accurate mode would achieve approximately 120 TFLOPS.

\subsection{GEMM Emulation Library}
\label{subsec:GEMM Emulation Library}
In this study, we developed a library that provides a complete emulation environment for GEMM operations based on the Ozaki-II scheme.
Details of the library and usage instructions are provided in~\cite{GEMMul8}.
The library supports both CUDA and HIP backends. The CUDA-based source code is automatically translated into HIP C++ when compiled under the HIP backend.
The implementation provides emulation routines for single- and double-precision GEMM, proposed in previous work~\cite{uchino_ozaki2}, as well as the CGEMM and ZGEMM emulation kernels developed in the present study.
The supported data types are 
\texttt{float}, 
\texttt{double}, 
\texttt{cuComplex},  
\texttt{cuDoubleComplex},  
\texttt{hipFloatComplex}, and
\texttt{hipDoubleComplex}.

To enable transparent integration with existing high-performance computing applications, the library can be activated by adding the path of the shared library to the environment variable \texttt{LD\_PRELOAD}. 
Once activated, the following GEMM calls to CUDA or HIP backends are dynamically redirected to the emulation routines, allowing users to evaluate GEMM emulation without modifying their applications:
\begin{itemize}
\item \texttt{cublasSgemm}, \texttt{cublasDgemm}
\item \texttt{cublasCgemm}, \texttt{cublasZgemm}
\item \texttt{cublasSgemm\_v2}, \texttt{cublasDgemm\_v2}
\item \texttt{cublasCgemm\_v2}, \texttt{cublasZgemm\_v2}
\item \texttt{cublasGemmEx}
\item \texttt{hipblasSgemm}, \texttt{hipblasDgemm}
\item \texttt{hipblasCgemm}, \texttt{hipblasZgemm}
\item \texttt{hipblasGemmEx}, \texttt{hipblasGemmEx\_v2}
\end{itemize}

The implementation consists of INT8 matrix multiplications, fixed-order basic arithmetic operations, and the evaluation of the $\log_{2}(\cdot)$ function in \eqref{eq:mu-fast}, \eqref{eq:nu-fast}, \eqref{eq:mu-accu}, and~\eqref{eq:nu-accu}.
Rounding errors arise only in the computation of the scaling vectors defined by~\eqref{eq:mu-fast}, \eqref{eq:nu-fast}, \eqref{eq:mu-accu}, and~\eqref{eq:nu-accu}, and in the accumulation and final reduction stages of the CRT.
Except for the logarithm evaluation, all computations are deterministic sequences of basic arithmetic, and thus the emulation produces bitwise-reproducible results as long as the precision of the $\log_{2}(\cdot)$ computation is identical.
When the API versions coincide, each backend employs identical implementations of \texttt{\_\_log2f}, ensuring reproducibility across runs.

\section{Numerical Results}
\label{sec:Numerical Results}
Here, we compare the following methods:
\begin{description}
    \item[Cgemm:] cublasCgemm or hipblasCgemm using native FP32
    \item[Zgemm:] cublasZgemm or hipblasZgemm using native FP64
    \item[Zgemm3m:] cublasZgemm3m using native FP64
    \item[fast-$N$:] Ozaki-II in fast mode with $N$ moduli
    \item[accu-$N$:] Ozaki-II in accurate mode with $N$ moduli
    \item[OS I-$S$:] Ozaki-I from cuBLAS in Fixed Mantissa Control mode with $S$ slices
\end{description}
hipBLAS does not provide hipblasCgemm3m or hipblasZgemm3m; the same limitation applies to rocBLAS, though it is not used in this study.
We empirically confirmed that both the performance and numerical accuracy of cublasCgemm3m are essentially identical to those of cublasCgemm; thus, we omit the former's results from the following comparisons.
The cuBLAS implementation of Ozaki-I-based emulation provides two execution modes: Dynamic Mantissa Control and Fixed Mantissa Control. 
In the Dynamic Mantissa Control mode, the routine determines the number of mantissa bits, or equivalently the number of slices, required for fixed-point emulation such that the numerical accuracy matches or exceeds that of FP64. 
If the required mantissa length exceeds either an internal limit or a user-specified maximum, the computation is automatically executed in native FP64. 
In the Fixed Mantissa Control mode, the user specifies the mantissa bit length used for the fixed-point representation. 
This mode reduces runtime overhead by eliminating dynamic precision selection, but it does not ensure FP64-level accuracy. 
In this study, the Fixed Mantissa Control mode is used to enable a direct comparison of computational throughput and to quantify the performance gain attributable to fixed-precision emulation.

\subsection{Accuracy}
\label{subsec:Accuracy}
Each element of the test matrices $A \in \mathbb{C}^{m \times k}$ and $B \in \mathbb{C}^{k \times n}$ was generated by independently assigning its real and imaginary parts as
\[
(\mathrm{rand}-0.5) \cdot \exp(\mathrm{randn} \cdot \phi),
\]
where $\mathrm{rand} \in (0,1]$ denotes a uniform random number and $\mathrm{randn}$ denotes a standard normal random number. 
Both types of random number were generated using either the cuRAND API or the hipRAND library. 
Following the methodology adopted in prior studies on Ozaki schemes~\cite{ozaki2012error,ozaki2013generalization,mukunoki2020,ootomo2024dgemm,uchino2025Performance,ozaki-scheme2,uchino_ozaki2}, we introduce the parameter $\phi$ to control the dynamic range of the input matrices.
In accordance with these studies, we specify $\phi \in \{0, 0.5, 1, 1.5\}$ for single-precision tests and $\phi \in \{0.5, 1, 2, 4\}$ for double-precision tests.
As $\phi$ increases, the maximum magnitude grows exponentially, whereas the median and interquartile range remain comparatively stable.
This confirms that $\phi$ effectively widens the dynamic range without substantially affecting the central tendency of the distribution.
Such controlled variation is essential for evaluating the performance and numerical stability of Ozaki schemes.

\begin{figure}[htbp]\centering
\includegraphics[width=\hsize]{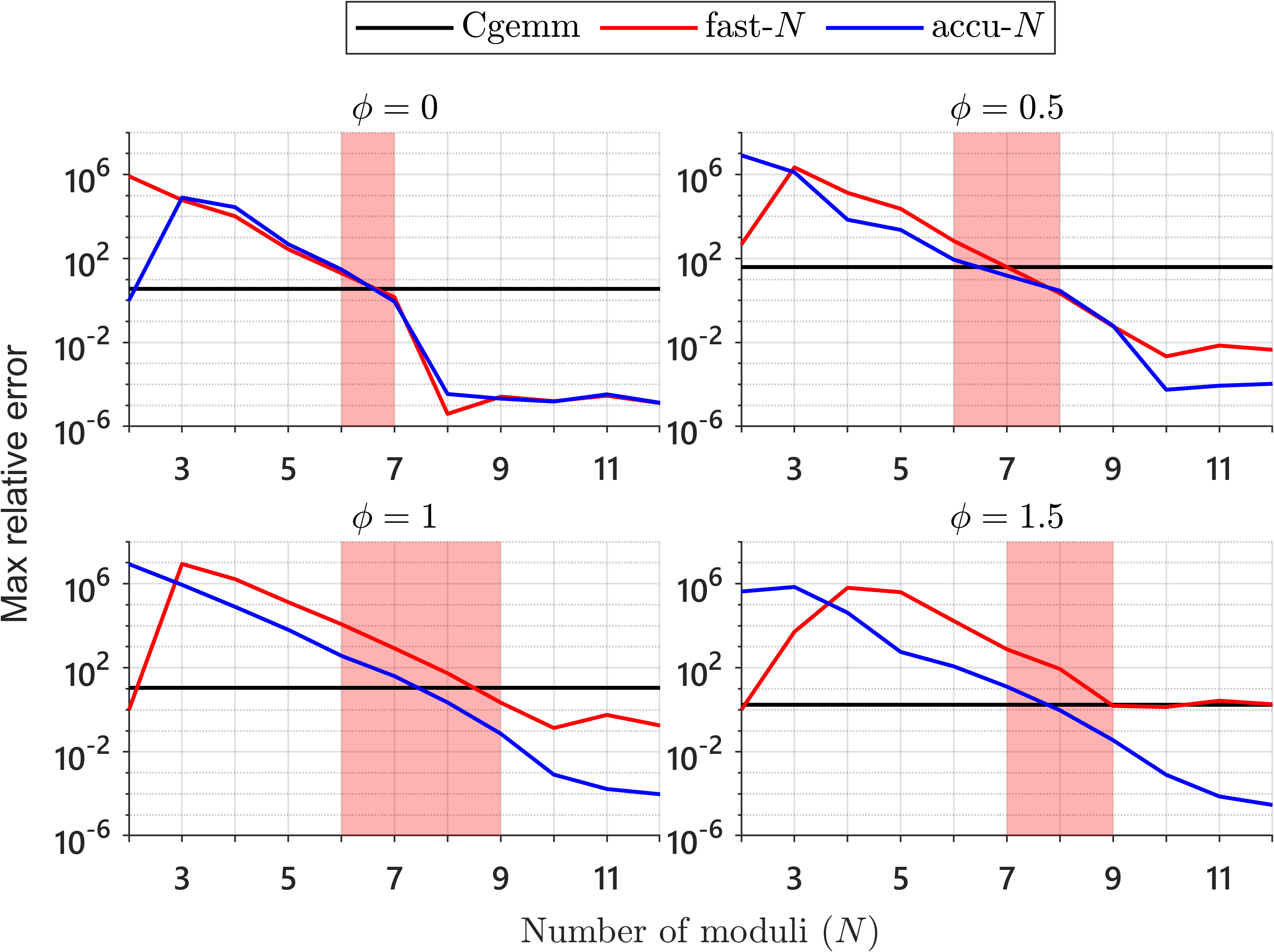}
\caption{Maximum relative error of single-precision complex matrix multiplication on NVIDIA GH200 Grace Hopper Superchip (CUDA Toolkit~13.0.88, gcc~11.5.0) for $m=n=1024$ and $k=16384$.}
\label{fig:accuracy-cgemm}
\end{figure}

\begin{figure}[htbp]\centering
\includegraphics[width=\hsize]{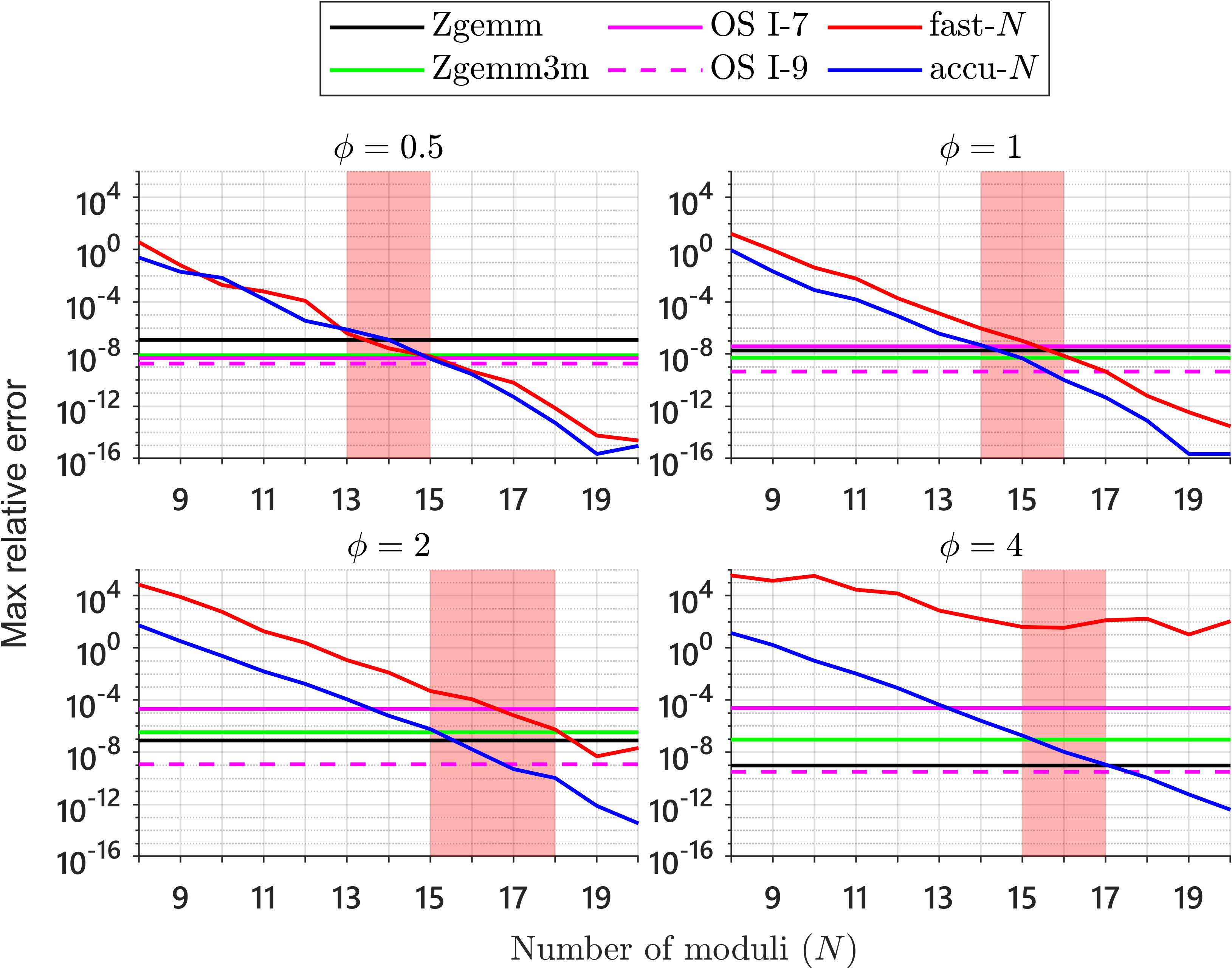}
\caption{Maximum relative error of double-precision complex matrix multiplication on NVIDIA GH200 Grace Hopper Superchip (CUDA Toolkit~13.0.88, gcc~11.5.0) for $m=n=1024$ and $k=16384$.}
\label{fig:accuracy-zgemm}
\end{figure}

Figs.~\ref{fig:accuracy-cgemm} and \ref{fig:accuracy-zgemm} show the accuracy of CGEMM and ZGEMM, respectively, observed on the GH200.
Similar results were obtained on other GPUs.
The maximum relative error in the figures was  calculated as
\[
\max_{i,j}\left( \max \left(  \frac{|\mathrm{Re}(\tilde{c}_{ij}) - \mathrm{Re}(c_{ij})|}{|\mathrm{Re}(c_{ij})|}, \frac{|\mathrm{Im}(\tilde{c}_{ij}) - \mathrm{Im}(c_{ij})|}{|\mathrm{Im}(c_{ij})|}  \right) \right),
\]
where $\tilde{C} \in \mathbb{C}^{m \times n}$ is the approximation of $AB$ obtained using the emulation methods and the standard complex GEMM routines, and $C \in \mathbb{C}^{m \times n}$ is the high-precision result of $AB$ computed using double-double arithmetic.

In CGEMM emulation, fast-$N$ with $6 \le N \le 9$ and accu-$N$ with $6 \le N \le 8$ provided accuracy comparable to that of the native CGEMM.
In ZGEMM emulation, fast-$N$ with $13 \le N \le 18$ and accu-$N$ with $13 \le N \le 17$ achieved accuracy comparable to that of the native ZGEMM.
As $\phi$ increases, the range of exponents for $A$ and $B$ expands, making it easier to overestimate the upper bound $2\sum_{h = 1}^k|a'_{ih}||b'_{hj}|$ in condition~\eqref{eq:cond}. 
DGEMM emulation in~\cite{uchino_ozaki2} required at least 14 moduli to achieve native DGEMM-level accuracy, whereas the proposed method attains native ZGEMM-level accuracy with only 13 moduli. 
This reduction is attributed to the use of the Karatsuba algorithm in the INT8 complex matrix multiplication, which suppresses precision loss more effectively than floating-point arithmetic.

The final reduction $\mathrm{mod}(s_{ij},P)$ of the CRT is defined as $s_{ij} - P \cdot \mathrm{round}(s_{ij} / P)$. 
This formulation entails subtracting nearly identical quantities, leading to loss of significance. 
Therefore, obtaining the target accuracy requires that both the modulo operation and the calculation of $s_{ij}$ be conducted at a precision level exceeding that of the final output.
Regarding implementation, the limiting accuracy is currently constrained by the arithmetic precision employed: CGEMM emulation relies on double-precision arithmetic, while ZGEMM emulation calculates $s_{ij}$ using the high-precision approach in~\eqref{eq:CRTaccumulation} and employs a simplified high-precision method based on double-double arithmetic for the modulo operation.
Enhancing the limiting accuracy therefore necessitates performing these operations with even higher precision.

\subsection{Performance}
\label{subsec:Performance}
\begin{figure}[htbp]\centering
\includegraphics[width=\hsize]{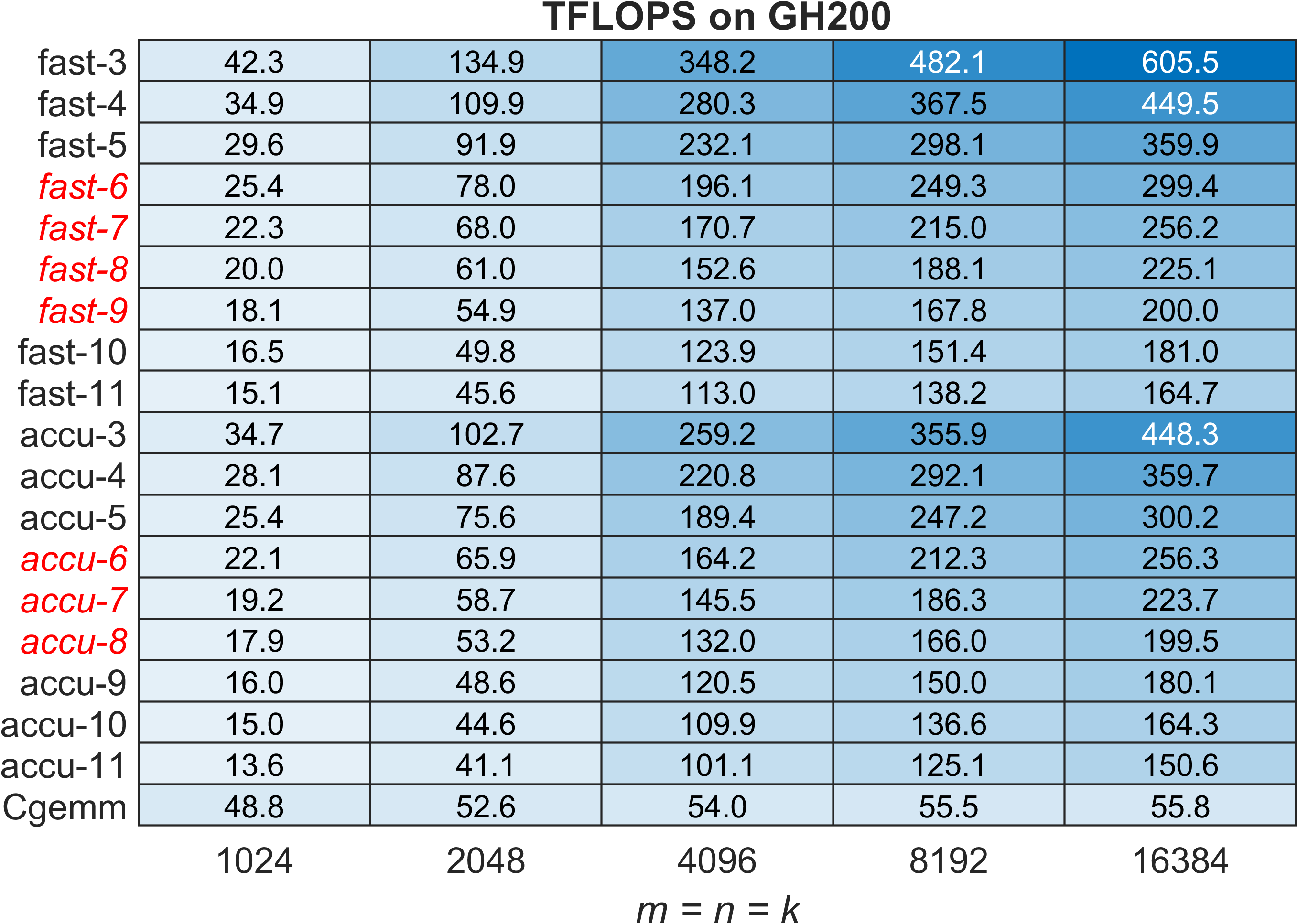}
\caption{Throughput performance (in TFLOPS) of single-precision complex matrix multiplication on NVIDIA GH200 Grace Hopper Superchip (CUDA Toolkit~13.0.88, gcc~11.5.0) for $m=k=n$. Methods shown in red and italics achieve CGEMM-level accuracy.}
\label{fig:tflops-cgemm-gh200}
\end{figure}

\begin{figure}[htbp]\centering
\includegraphics[width=\hsize]{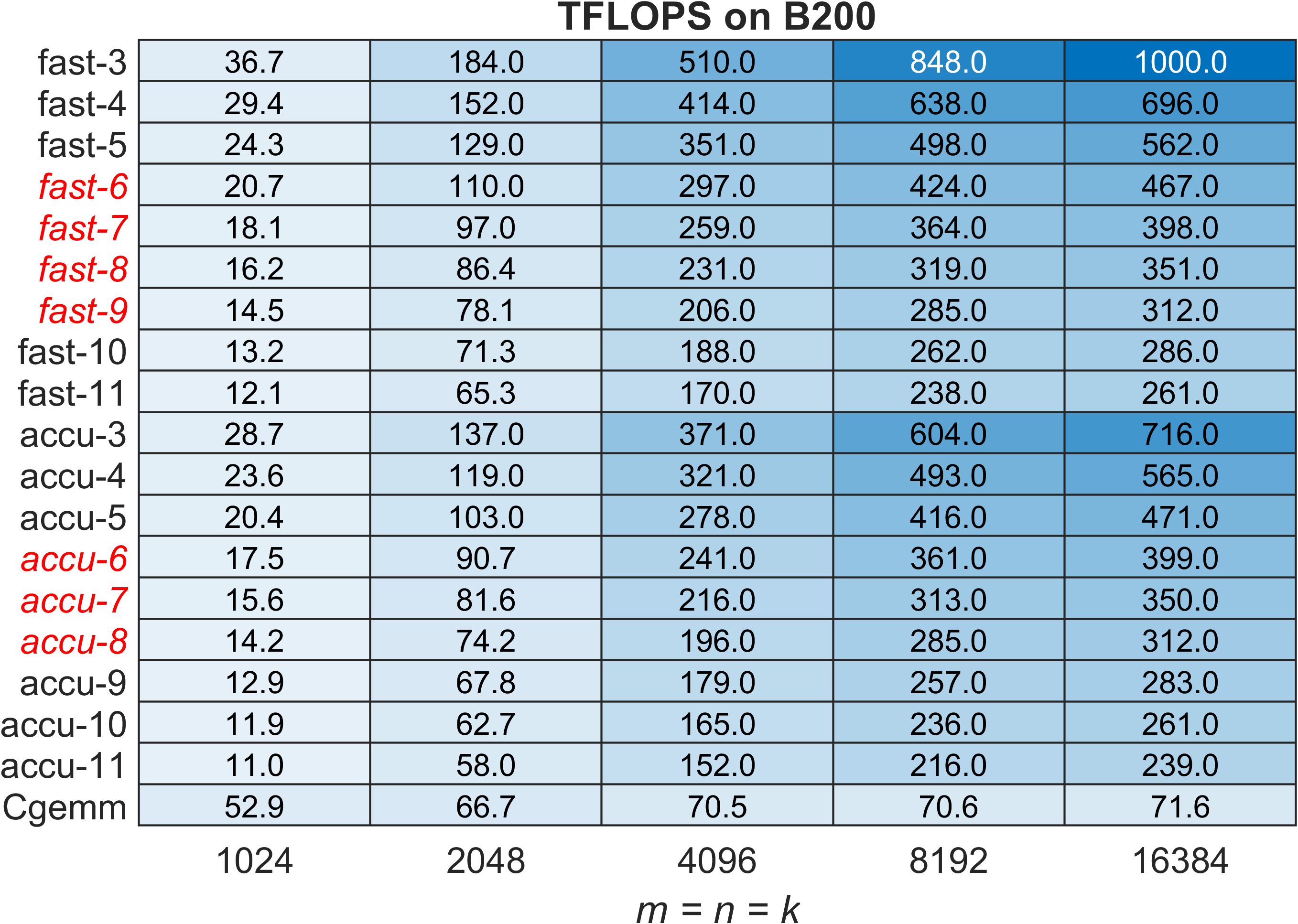}
\caption{Throughput performance (in TFLOPS) of single-precision complex matrix multiplication on NVIDIA B200 Blackwell GPU with Intel Xeon Platinum 8570 CPU (CUDA Toolkit~12.8.93, g++~13.3.0) for $m=k=n$. Methods shown in red and italics achieve CGEMM-level accuracy.}
\label{fig:tflops-cgemm-b200}
\end{figure}

\begin{figure}[htbp]\centering
\includegraphics[width=\hsize]{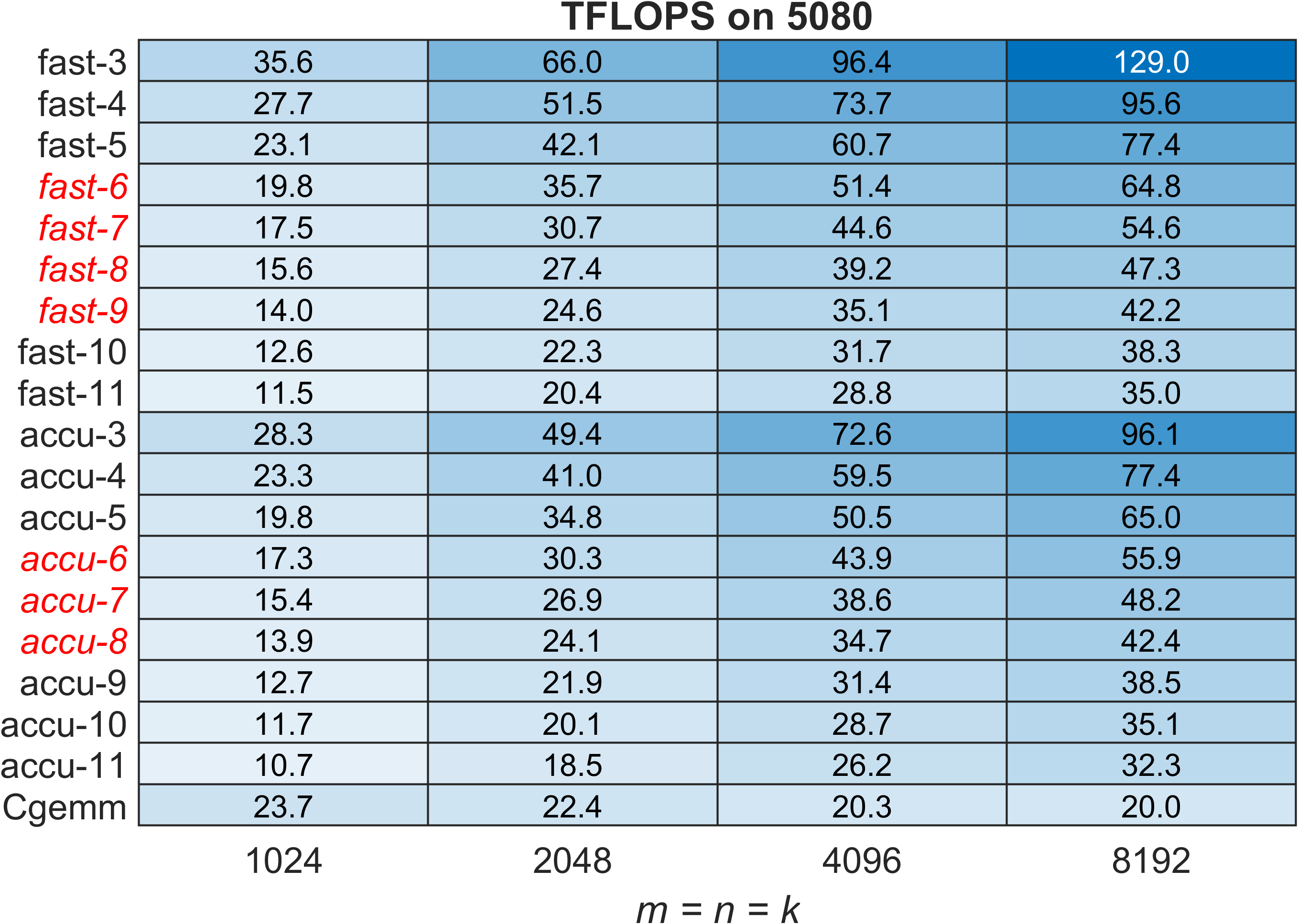}
\caption{Throughput performance (in TFLOPS) of single-precision complex matrix multiplication on NVIDIA GeForce RTX 5080 GPU with AMD Ryzen 7 3700X CPU (CUDA Toolkit~13.0.88, gcc~11.5.0) for $m=k=n$. Methods shown in red and italics achieve CGEMM-level accuracy.}
\label{fig:tflops-cgemm-5080}
\end{figure}

\begin{figure}[htbp]\centering
\includegraphics[width=\hsize]{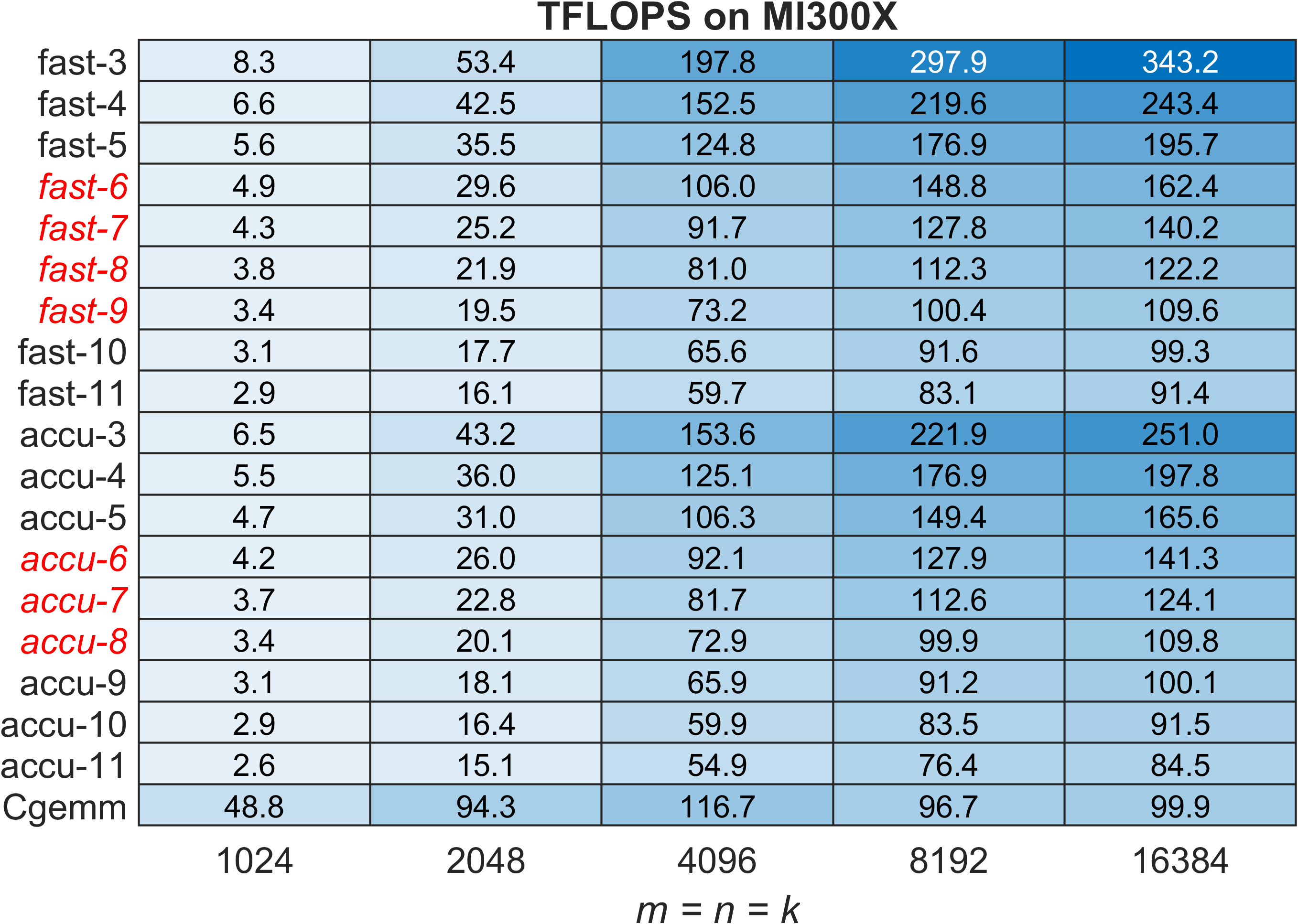}
\caption{Throughput performance (in TFLOPS) of single-precision complex matrix multiplication on AMD Instinct MI300X GPU with AMD EPYC 9534 CPU (ROCm 6.4.2, clang~19.0.0git) for $m=k=n$. Methods shown in red and italics achieve CGEMM-level accuracy.}
\label{fig:tflops-cgemm-mi300x}
\end{figure}

\begin{figure}[htbp]\centering
\includegraphics[width=\hsize]{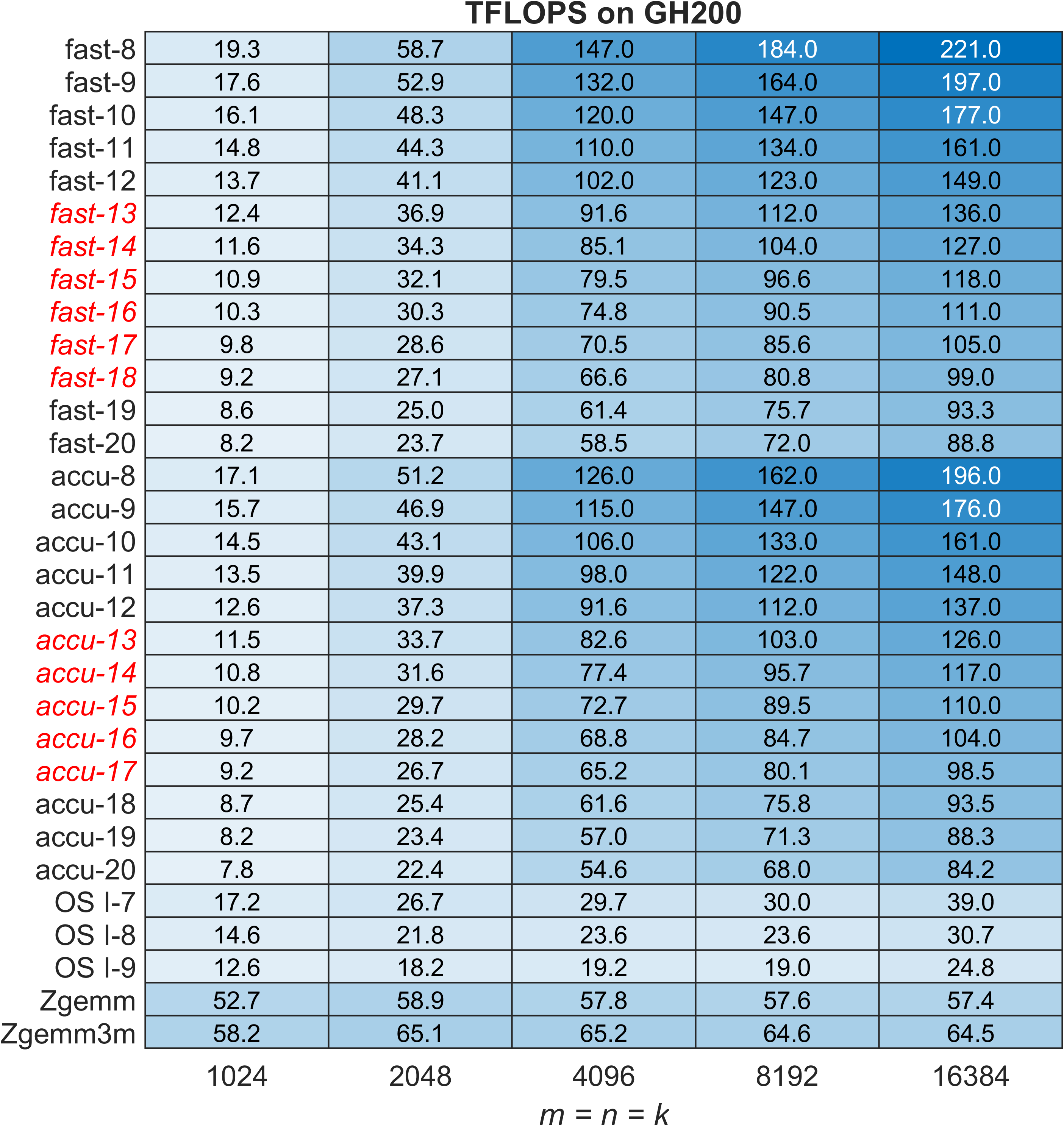}
\caption{Throughput performance (in TFLOPS) of double-precision complex matrix multiplication on NVIDIA GH200 Grace Hopper Superchip (CUDA Toolkit~13.0.88, gcc~11.5.0) for $m=k=n$. Methods shown in red and italics achieve ZGEMM-level accuracy.}
\label{fig:tflops-zgemm-gh200}
\end{figure}

\begin{figure}[htbp]\centering
\includegraphics[width=\hsize]{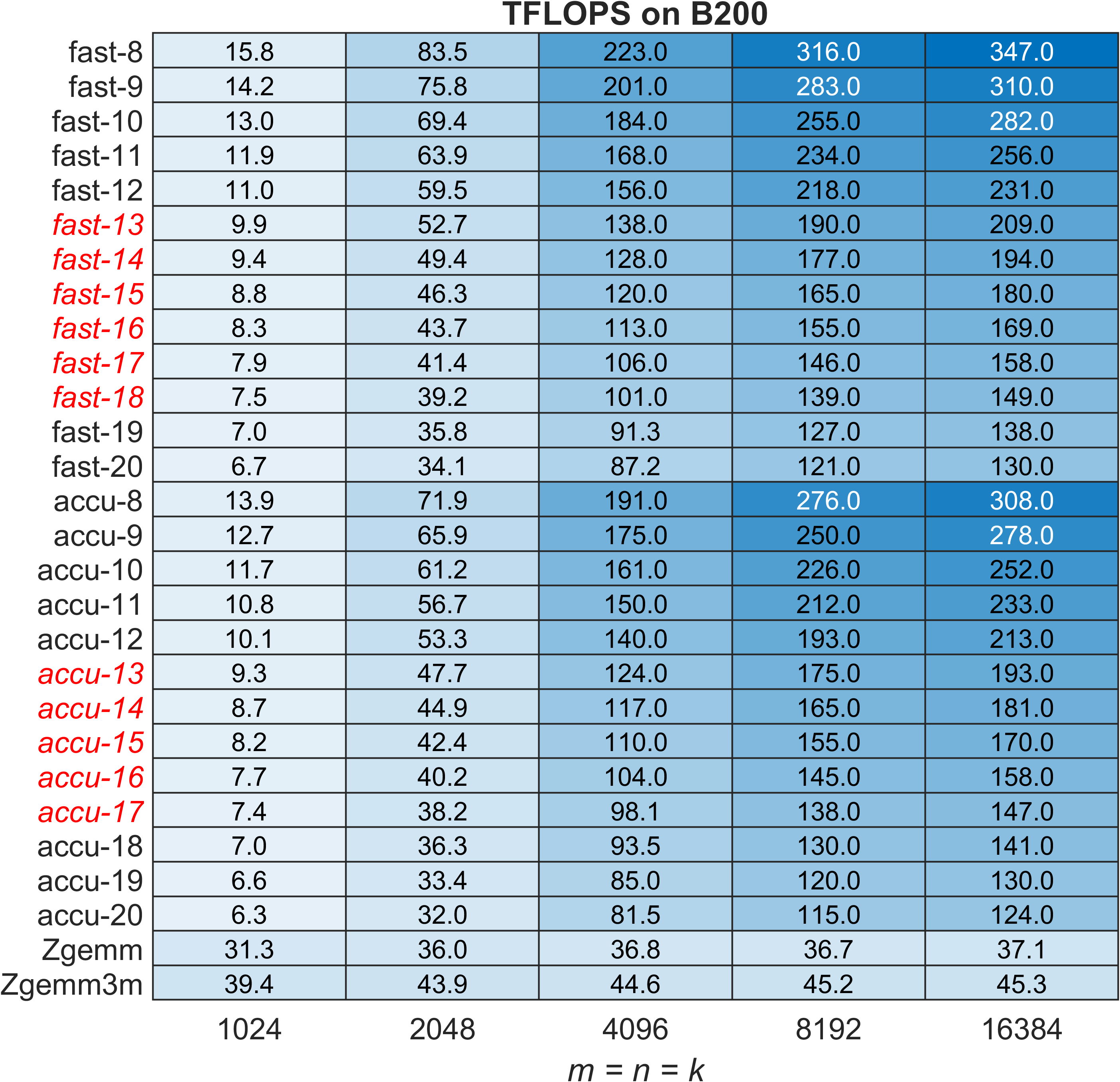}
\caption{Throughput performance (in TFLOPS) of double-precision complex matrix multiplication on NVIDIA B200 Blackwell GPU with Intel Xeon Platinum 8570 CPU (CUDA Toolkit~12.8.93, g++~13.3.0) for $m=k=n$. Methods shown in red and italics achieve ZGEMM-level accuracy.}
\label{fig:tflops-zgemm-b200}
\end{figure}

\begin{figure}[htbp]\centering
\includegraphics[width=\hsize]{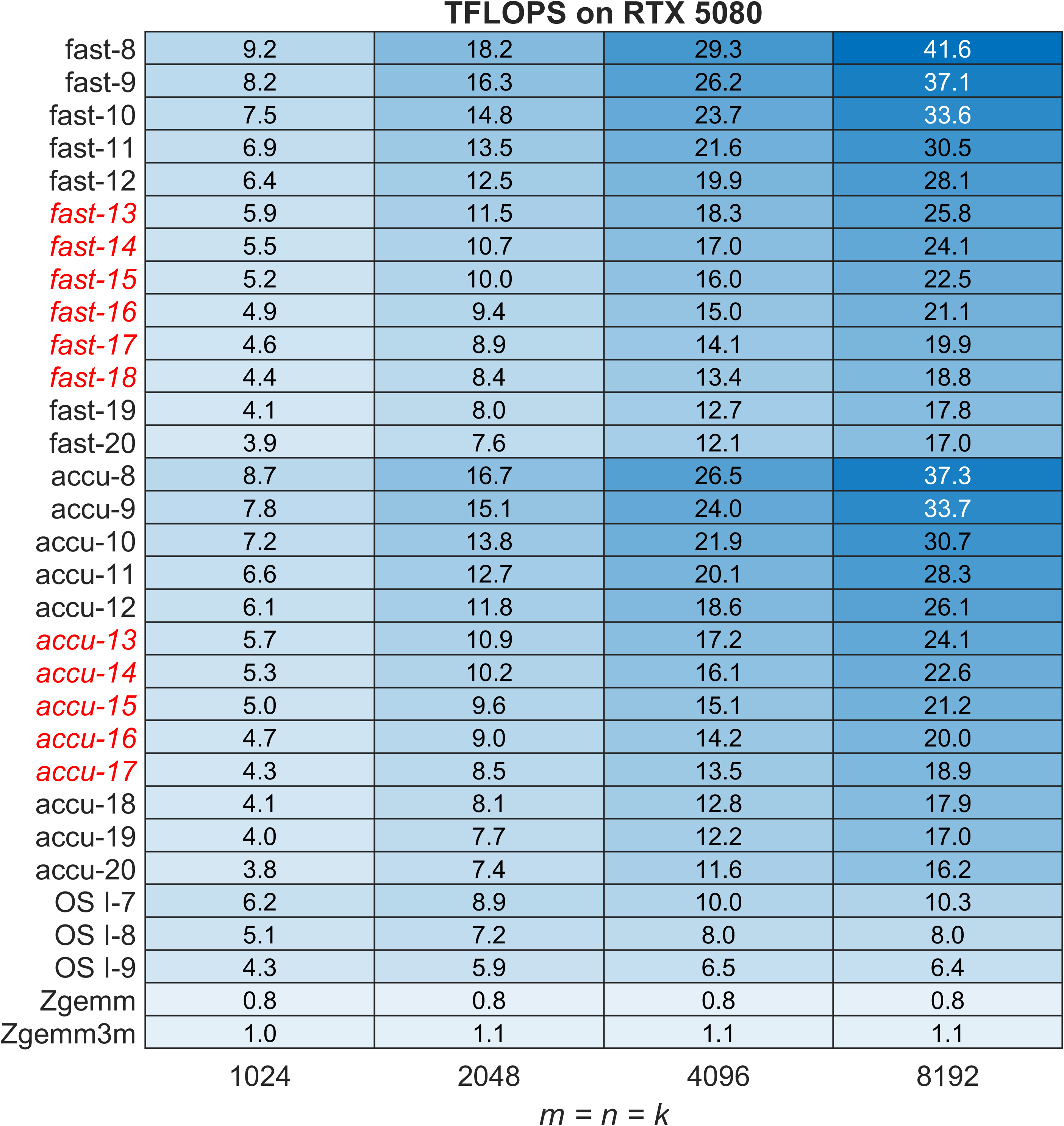}
\caption{Throughput performance (in TFLOPS) of double-precision complex matrix multiplication on NVIDIA GeForce RTX 5080 GPU with AMD Ryzen 7 3700X CPU (CUDA Toolkit~13.0.88, gcc~11.5.0) for $m=k=n$. Methods shown in red and italics achieve ZGEMM-level accuracy.}
\label{fig:tflops-zgemm-5080}
\end{figure}

\begin{figure}[htbp]\centering
\includegraphics[width=\hsize]{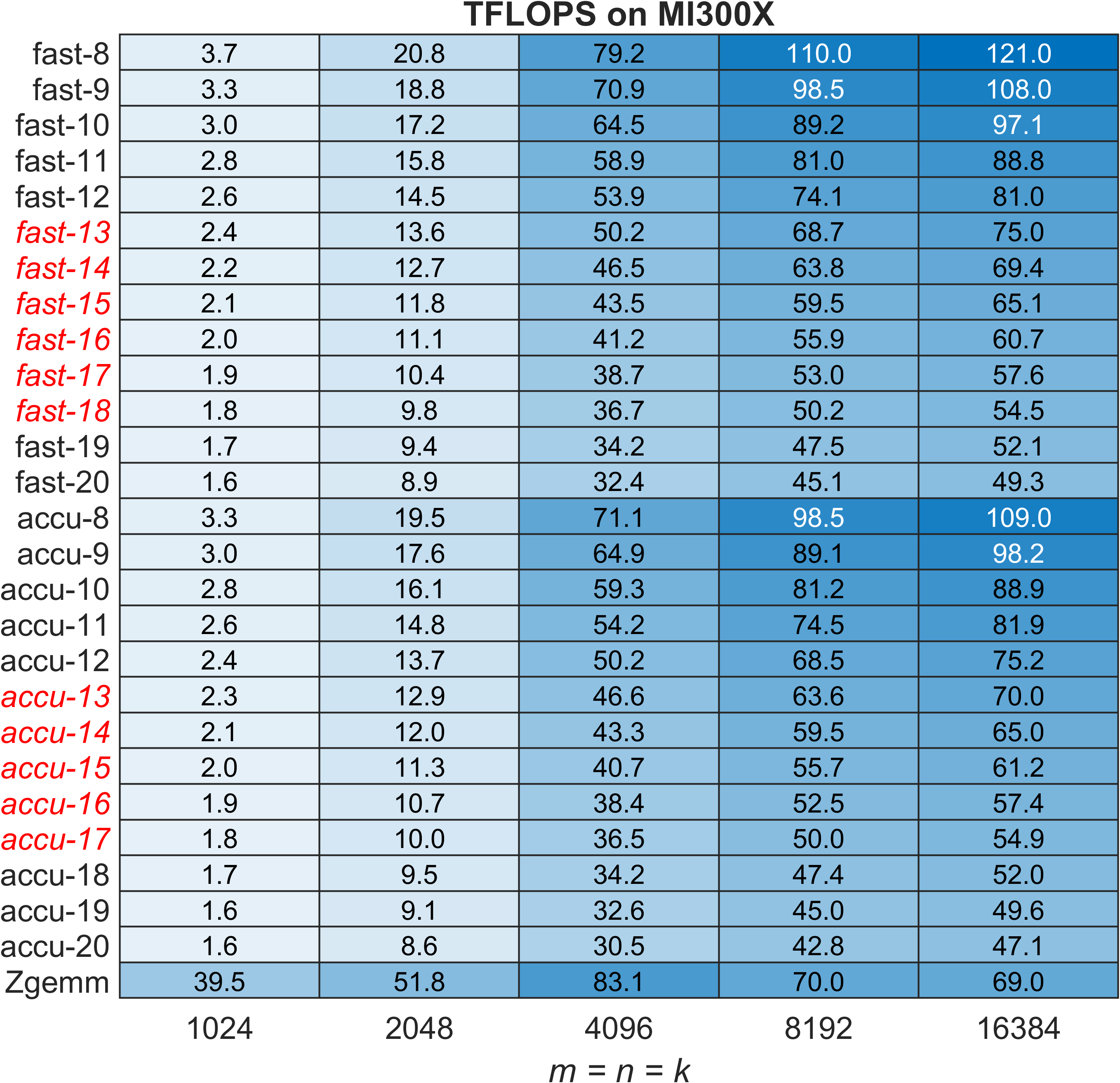}
\caption{Throughput performance (in TFLOPS) of double-precision complex matrix multiplication on AMD Instinct MI300X GPU with AMD EPYC 9534 CPU (ROCm 6.4.2, clang~19.0.0git) for $m=k=n$. Methods shown in red and italics achieve ZGEMM-level accuracy.}
\label{fig:tflops-zgemm-mi300x}
\end{figure}

\begin{figure}[htbp]\centering
\includegraphics[width=\hsize]{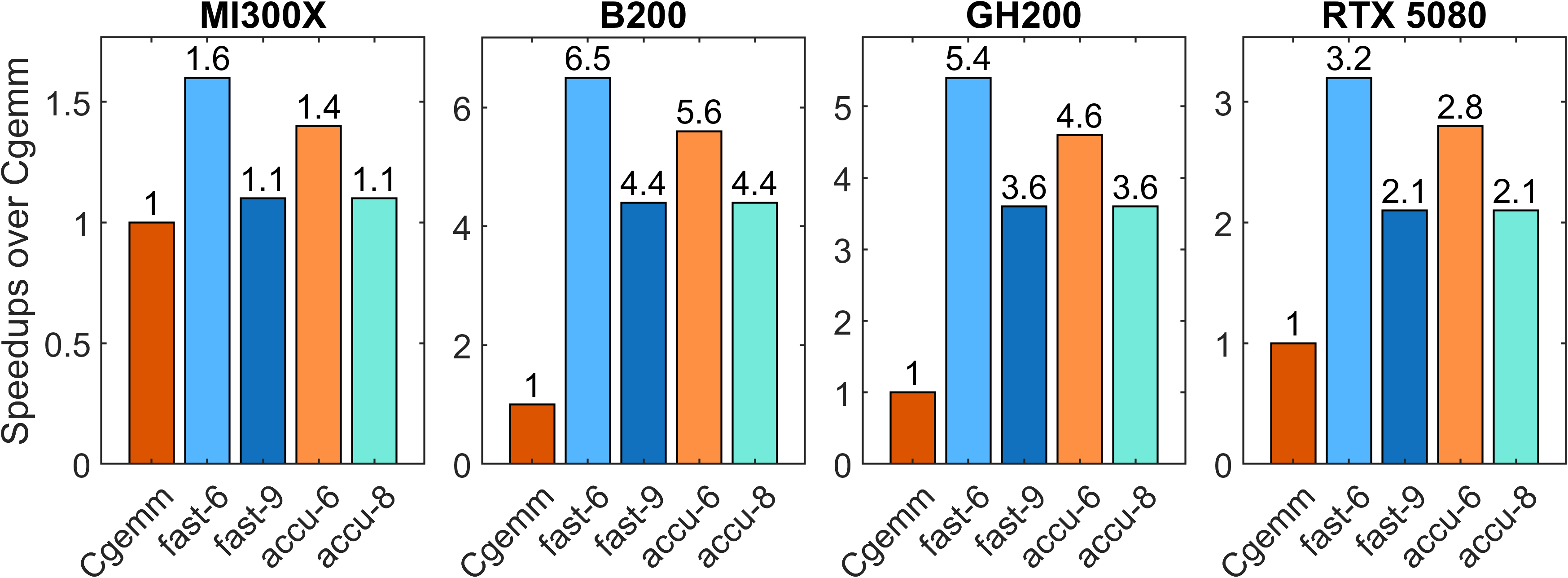}
\caption{Speedups of single-precision complex matrix multiplication over native CGEMM for $m=k=n=16384$}
\label{fig:speedups_c}
\end{figure}

\begin{figure}[htbp]\centering
\includegraphics[width=\hsize]{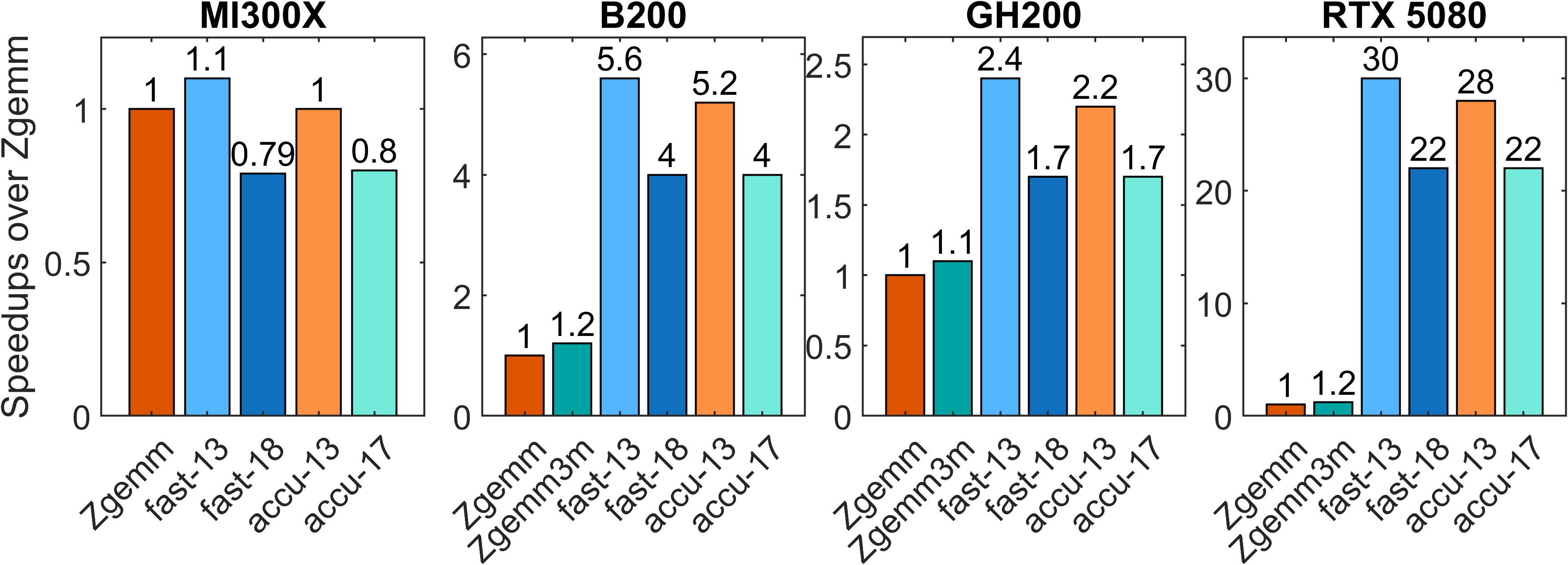}
\caption{Speedups of double-precision complex matrix multiplication over native ZGEMM for $m=k=n=16384$}
\label{fig:speedups_z}
\end{figure}

Figs.~\ref{fig:tflops-cgemm-gh200}--\ref{fig:tflops-zgemm-mi300x} show the throughput performance of CGEMM and ZGEMM.
Figures~\ref{fig:speedups_c} and~\ref{fig:speedups_z} show the speedups of each method relative to the native CGEMM and ZGEMM.
For CGEMM, fast-$N$ with $6 \le N \le 9$ and accu-$N$ with $6 \le N \le 8$ achieved substantial speedups on the GH200 and B200, reaching approximately 3.6--5.4$\times$ and 4.4--6.5$\times$, respectively.
On the MI300X, however, the improvements were modest, with gains limited to 1.1--1.6$\times$.
For ZGEMM, where fast-$N$ spans $13 \le N \le 18$ and accu-$N$ spans  $13 \le N \le 17$, a similar trend was observed.
The GH200 delivered moderate acceleration, achieving 1.7--2.4$\times$ speedups, whereas the B200 consistently attained the highest speedups (4.0--5.6$\times$).
In contrast, the MI300X again exhibited only minimal improvement, ranging from 0.8$\times$ to 1.1$\times$, effectively matching or slightly underperforming the baseline routine.
In addition, on the RTX 5080, for CGEMM, the proposed methods yielded 2.1--3.2$\times$ speedups and for ZGEMM, the proposed methods yielded 22--30$\times$ speedups.
Overall, the Ozaki-II emulation yielded significant performance benefits on NVIDIA architectures, especially on the B200 and RTX 5080, while demonstrating only limited impact on the MI300X.

These performance characteristics correlate strongly with the architectural balance of arithmetic throughput on each GPU.
The GH200 provides a ratio of $\text{FP64} : \text{FP32} : \text{INT8} = 1 : 1 : 29.5$, indicating that INT8 units are more abundant than FP64 and FP32 units.
The B200 exhibits an even more pronounced imbalance, $1 : 2 : 121.6$, where INT8 throughput overwhelmingly dominates the floating-point pipelines.
This architectural skew directly amplifies the benefit of the proposed kernels, which replace large portions of floating-point arithmetic with INT8-intensive computation.
In contrast, the MI300X offers a more moderate ratio of $1 : 1 : 16$, with INT8 throughput being moderately higher than floating-point throughput.
As a result, the computational substitution enabled by our method yields only marginal acceleration on the MI300X, consistent with the limited speedups observed in the benchmarks.
For the RTX 5080, the observed performance gains, especially the very large speedups in ZGEMM, suggest a much higher INT8 throughput relative to that of floating-point units. The architectural balance on the RTX 5080 corresponds to $\text{FP64} : \text{FP32} : \text{INT8} = 1 : 64 : 512$, which explains the pronounced acceleration when using INT8-based emulation.

Compared to the Ozaki-I emulation implemented in cuBLAS, the proposed methods delivered substantially higher throughput on both the GH200 and RTX 5080.
On the GH200, Ozaki-I with 7--9 slices achieved 25--39 TFLOPS, whereas the proposed methods achieved 99--136 TFLOPS.
On the RTX 5080, Ozaki-I achieved 6.4--10 TFLOPS, whereas the proposed methods achieved 19--26 TFLOPS.
These disparities arise from fundamental algorithmic differences: Ozaki-I with $S$ slices requires $S(S+1)/2$ INT8 complex matrix multiplications, whereas the proposed technique with $N$ moduli requires only $N$ INT8 complex matrix multiplications, whose cost is further reduced by a factor of $3/4$ using the Karatsuba method.
This reduction in both the number and cost of INT8 matrix multiplication kernel invocations explains the substantially higher performance of the proposed methods.

\subsection{Supplemental Comparison of Real-Valued Emulation}
\label{subsec:Supplemental Comparison for Real-Valued Emulation}
In this section, we focus on double-precision real matrix multiplication.
We compare two emulation strategies: the Ozaki-I based emulation provided by cuBLAS and the Ozaki-II based emulation proposed in~\cite{uchino_ozaki2}, enhanced with the blocking technique introduced in Section~\ref{sec:Proposed Methods}.
This supplemental analysis aims to isolate the behavior of Ozaki-type schemes in the real domain and to clarify the performance implications of incorporating blocking into Ozaki-II under conditions consistent with real-valued DGEMM emulation.

According to~\cite{uchino_ozaki2}, the fast mode and the accurate mode require 14--18 and 14--17 moduli, respectively, to achieve DGEMM-level accuracy.
On the GH200 with matrix size $m=n=k=16384$, the Ozaki-II fast-$N$ configuration with $14 \le N \le 18$ achieved 63--82 TFLOPS and the accu-$N$ configuration with $14 \le N \le 17$ achieved 63--75 TFLOPS.
Incorporating the blocking technique improved performance across all settings: the blocked fast-$N$ version achieved 72--93 TFLOPS and the blocked accu-$N$ version achieved 71--85 TFLOPS.
cuBLAS's Ozaki-I implementation with 7--10 slices achieved only 20--39 TFLOPS and the standard DGEMM kernel achieved 61 TFLOPS.

On the RTX 5080 with $m=n=k=8192$, the fast-$N$ and accu-$N$ configurations of the proposed methods achieved 12--16 TFLOPS and 12--15 TFLOPS, respectively, and cuBLAS's Ozaki-I achieved 5.4--10 TFLOPS.
In comparison, the native DGEMM kernel achieved 0.84 TFLOPS, highlighting the substantial advantage of INT8-intensive Ozaki-type emulation on this consumer architecture.

These results demonstrate that blocking substantially enhances the performance of Ozaki-II for large-scale problems, particularly on bandwidth-rich GPUs such as the B200, and that Ozaki-II, whether blocked or unblocked, consistently outperforms cuBLAS's Ozaki-I due to its reduced number of INT8 matrix multiplication invocations.

\section{Conclusion}
\label{sec:Conclusion}
This study proposed emulation methods for complex matrix multiplication based on the Ozaki-II scheme and developed a portable library that supports execution on both AMD and NVIDIA GPUs.
The numerical results demonstrated that the proposed methods outperform the standard GEMM routines in cuBLAS and hipBLAS, which are implemented using native FP32 and FP64 arithmetic.
In addition, the proposed methods outperform emulation using the Ozaki-I scheme implemented in cuBLAS.

Future work will include comparisons with the BF16x9 CGEMM emulation algorithm provided for NVIDIA B200 and B300 GPUs introduced in cuBLAS 12.9. 
A major limitation of the proposed methods, as well as emulation-based approaches in general, is the substantial working memory required. 
This overhead is currently unavoidable when aiming to achieve high performance without relying on FP32 or FP64 matrix multiplication. 
Addressing this issue remains an open challenge not only for emulation techniques but also for high-performance computing applications that employ them.

The Ozaki-I and Ozaki-II schemes utilize INT8 matrix engines, benefiting from the fact that the error-free low-precision matrix multiplications in these schemes do not require exponent handling, enabling efficient synergy with INT8 operations with INT32 accumulation. 
When using FP16 or BF16 matrix engines with FP32 accumulation, the effective representable integer range is limited to 24 bits due to the FP32 significand, leading to significantly reduced computational efficiency compared to that of INT32 accumulation. 
However, some recent GPUs, such as the NVIDIA B300, provide reduced INT8 capabilities. 
Support for such hardware architectures will be considered in future work.

\section*{Acknowledgment}
We thank Prof. Rio Yokota for his cooperation in conducting the experiments on the B200.
This study was supported by Japan Society for the Promotion of Science Grant-in-Aid Numbers 25K03126 (Scientific Research B) and 24K23874 (Research Activity Start-up).

\bibliographystyle{IEEEtran}
\bibliography{refs}

@InProceedings{mukunoki2020,
author="Mukunoki, Daichi
and Ozaki, Katsuhisa
and Ogita, Takeshi
and Imamura, Toshiyuki",
editor="Sadayappan, Ponnuswamy
and Chamberlain, Bradford L.
and Juckeland, Guido
and Ltaief, Hatem",
title="DGEMM Using Tensor Cores, and Its Accurate and Reproducible Versions",
booktitle="High Performance Computing",
year="2020",
publisher="Springer International Publishing",
address="Cham",
pages="230--248",
}

@online{tensorcore,
author = {{NVIDIA Corporation}},
title = {{NVIDIA Tensor Cores}},
year = {2025},
url = {https://www.nvidia.com/en-us/data-center/tensor-cores/},
note = {retrieved 25 July, 2025}
}

@online{matrixcore,
author = {{Advanced Micro Devices, Inc.}},
title = {{AMD matrix cores}},
year = {2025},
url={https://rocm.blogs.amd.com/software-tools-optimization/matrix-cores/README.html},
note = {retrieved 21 March, 2025}
}

@online{intelXecore,
author = {Ramesh Perumal and Nikitha Chinthalapani and Rohit D’Souza},
title = {Heterogeneous AI Powerhouse: Unveiling the Hardware and Software Foundation of Intel\textsuperscript{\textregistered} Core\textsuperscript{\texttrademark} Ultra Processors for the Edge, White Paper},
year = {2024},
url={https://www.intel.com/content/www/us/en/content-details/817736}
}

@online{tpu,
author = {{Google LLC}},
title = {{Cloud Tensor Processing Unit}},
year = {2025},
url = {https://cloud.google.com/tpu/docs/intro-to-tpu},
note = {retrieved 21 March, 2025}
}

@online{copilot+pc,
author = {{Microsoft}},
title = {{Copilot+ PCs}},
year = {2025},
url={https://www.microsoft.com/en-us/windows/copilot-plus-pcs},
note = {retrieved 27 November, 2025}
}

@online{SME,
author = {Zenon Xiu},
title = {Part 1: Arm Scalable Matrix Extension (SME) Introduction},
year = {2025},
url={https://developer.arm.com/community/arm-community-blogs/b/architectures-and-processors-blog/posts/arm-scalable-matrix-extension-introduction},
note = {retrieved 27 November, 2025}
}

@article{ozaki2012error,
	title={Error-free transformations of matrix multiplication by using fast routines of matrix multiplication and its applications},
	author={Ozaki, Katsuhisa and Ogita, Takeshi and Oishi, Shin'ichi and Rump, Siegfried M.},
	journal={Numerical Algorithms},
	volume={59},
	number={1},
	pages={95--118},
	year={2012},
	publisher={Springer},
doi = {https://doi.org/10.1007/s11075-011-9478-1}
}

@article{ozaki2013generalization,
  title={Generalization of error-free transformation for matrix multiplication and its application},
  author={Ozaki, Katsuhisa and Ogita, Takeshi and Oishi, Shin'ichi and Rump, Siegfried M},
  journal={Nonlinear Theory and Its Applications, IEICE},
  volume={4},
  number={1},
  pages={2--11},
  year={2013},
doi = {https://doi.org/10.1587/nolta.4.2},
  publisher={The Institute of Electronics, Information and Communication Engineers}
}

@article{ootomo2024dgemm,
author = {Hiroyuki Ootomo and Katsuhisa Ozaki and Rio Yokota},
title ={DGEMM on integer matrix multiplication unit},
journal = {The International Journal of High Performance Computing Applications},
volume = {38},
number = {4},
pages = {297-313},
year = {2024},
doi = {10.1177/10943420241239588},
}

@article{uchino2025Performance,
author = {Yuki Uchino and Katsuhisa Ozaki and Toshiyuki Imamura},
title ={Performance enhancement of the Ozaki Scheme on integer matrix multiplication unit},
journal = {The International Journal of High Performance Computing Applications},
volume = {39},
number = {3},
pages = {462--476},
year = {2025},
doi = {10.1177/10943420241313064},
}

@misc{ozaki-scheme2,
      title={{Ozaki Scheme II}: A {GEMM}-oriented emulation of floating-point matrix multiplication using an integer modular technique}, 
      author={Katsuhisa Ozaki and Yuki Uchino and Toshiyuki Imamura},
      year={2025},
      eprint={2504.08009},
      archivePrefix={arXiv},
      primaryClass={cs.MS},
}

@inproceedings{uchino_ozaki2,
    author = {Uchino, Yuki and Ozaki, Katsuhisa and Imamura, Toshiyuki},
    title = {High-Performance and Power-Efficient Emulation of Matrix Multiplication using INT8 Matrix Engines},
    year = {2025},
    isbn = {9798400718717},
    publisher = {Association for Computing Machinery},
    address = {St. louis, MO, USA},
    url = {https://doi.org/10.1145/3731599.3767539},
    doi = {10.1145/3731599.3767539},
    booktitle = {Proceedings of the SC '25 Workshops of the International Conference for High Performance Computing, Networking, Storage and Analysis},
    pages = {1824-1831},
    numpages = {8},
    series = {SC Workshops '25}
}

@software{GEMMul8,
    title = {{GEMMul8}: {GEMM} emulation using int8 matrix engines based on the {Ozaki scheme II}},
    year={2025},
    author={Yuki Uchino},
    license = {{MIT}},
    institution = {{RIKEN Center for Computational Science}},
    howpublished = {{R-CCS GitHub} repository},
    url = {https://github.com/RIKEN-RCCS/GEMMul8},
    repository= {https://github.com/RIKEN-RCCS/GEMMul8},
}

@misc{mukunoki2025dgemmfp64arithmetic,
      title={DGEMM without FP64 Arithmetic - Using FP64 Emulation and FP8 Tensor Cores with Ozaki Scheme}, 
      author={Daichi Mukunoki},
      year={2025},
      eprint={2508.00441},
      archivePrefix={arXiv},
      primaryClass={cs.PF},
      url={https://arxiv.org/abs/2508.00441}, 
}

@misc{cuda,
    author = {{NVIDIA Corporation}},
    title = {{CUDA C++ Programming Guide Release 13.0}},
    year = {2025},
    url = {https://docs.nvidia.com/cuda/cuda-c-programming-guide},
    note = {retrieved 7 October, 2025}
}

@techreport{muller:inria-00070503,
  TITLE = {{On the definition of ulp(x)}},
  AUTHOR = {Muller, Jean-Michel},
  URL = {https://inria.hal.science/inria-00070503},
  TYPE = {Research Report},
  NUMBER = {RR-5504, LIP RR-2005-09},
  PAGES = {16},
  INSTITUTION = {{INRIA, LIP}},
  YEAR = {2005},
  MONTH = Feb,
}

@misc{hip,
    author = {{Advanced Micro Devices, Inc.}},
    url = {https://rocm.docs.amd.com/projects/HIP/en/latest/reference/math_api.html},
    title = {{HIP math API --- HIP 7.0.51831 Documentation}},
    year = {2025},
    note = {retrieved 7 October, 2025},
}

@INPROCEEDINGS{11196413,
  author={Dongarra, Jack and Gunnels, John and Bayraktar, Harun and Haidar, Azzam and Ernst, Dan},
  booktitle={2025 IEEE High Performance Extreme Computing Conference (HPEC)}, 
  title={Accelerating Supercomputing: AI-Hardware-Driven Innovation for Speed and Efficiency}, 
  year={2025},
  volume={},
  number={},
  pages={1-7},
  doi={10.1109/HPEC67600.2025.11196413}}

@INPROCEEDINGS{11196192,
  author={Luszczek, Piotr and Gadepally, Vijay and Anderson, LaToya and Arcand, William and Bestor, David and Bergeron, William and Bonn, Alex and Burrill, Daniel J. and Byun, Chansup and Houle, Michael and Hubbell, Matthew and Jones, Michael and Michaleas, Peter and Morales, Guillermo and Mullen, Julia and Prout, Andrew and Reuther, Albert and Rosa, Antonio and Yee, Charles and Kepner, Jeremy},
  booktitle={2025 IEEE High Performance Extreme Computing Conference (HPEC)}, 
  title={Performance and Numerical Aspects of Decompositional Factorizations with FP64 Floating-Point Emulation in INT8}, 
  year={2025},
  volume={},
  number={},
  pages={1-7},
  doi={10.1109/HPEC67600.2025.11196192}}

@misc{dongarra2024hardwaretrendsimpactingfloatingpoint,
      title={Hardware Trends Impacting Floating-Point Computations In Scientific Applications}, 
      author={Jack Dongarra and John Gunnels and Harun Bayraktar and Azzam Haidar and Dan Ernst},
      year={2024},
      eprint={2411.12090},
      archivePrefix={arXiv},
      primaryClass={math.NA},
      url={https://arxiv.org/abs/2411.12090}, 
}

@misc{mukunoki2025sparseiterativesolversusing,
      title={Sparse Iterative Solvers Using High-Precision Arithmetic with Quasi Multi-Word Algorithms}, 
      author={Daichi Mukunoki and Katsuhisa Ozaki},
      year={2025},
      eprint={2510.13536},
      archivePrefix={arXiv},
      primaryClass={cs.MS},
      url={https://arxiv.org/abs/2510.13536}, 
}

@misc{schwarz2025guaranteed,
  title={Guaranteed DGEMM Accuracy While Using Reduced Precision Tensor Cores Through Extensions of the Ozaki Scheme},
  author={Schwarz, Angelika and Anders, Anton and Brower, Cole and Bayraktar, Harun and Gunnels, John and Clark, Kate and Xu, RuQing G and Rodriguez, Samuel and Cayrols, Sebastien and Tabaszewski, Pawe{\l} and Podlozhnyuk, Victor},
  year={2025},
  archivePrefix={arXiv},
  primaryClass={cs.DC},
  url={https://arxiv.org/abs/2511.13778}, 
}

@misc{abdelfattah2025analysisfloatingpointmatrixmultiplication,
      title={Analysis of Floating-Point Matrix Multiplication Computed via Integer Arithmetic}, 
      author={Ahmad Abdelfattah and Jack Dongarra and Massimiliano Fasi and Mantas Mikaitis and Françoise Tisseur},
      year={2025},
      eprint={2506.11277},
      archivePrefix={arXiv},
      primaryClass={math.NA},
      url={https://arxiv.org/abs/2506.11277}, 
}

@misc{brower2025mixedprecisionabinitiotensor,
      title={Mixed-precision ab initio tensor network state methods adapted for NVIDIA Blackwell technology via emulated FP64 arithmetic}, 
      author={Cole Brower and Samuel Rodriguez Bernabeu and Jeff Hammond and John Gunnels and Sotiris S. Xanthea and Martin Ganahl and Andor Menczer and Örs Legeza},
      year={2025},
      eprint={2510.04795},
      archivePrefix={arXiv},
      primaryClass={physics.chem-ph},
      url={https://arxiv.org/abs/2510.04795}, 
}

@online{MI300A,
author = {{Advanced Micro Devices, Inc.}},
title = {AMD INSTINCT MI300X APU},
year = {2025},
url={https://www.amd.com/content/dam/amd/en/documents/instinct-tech-docs/data-sheets/amd-instinct-mi300a-data-sheet.pdf},
note = {retrieved 5 December, 2025}
}

@online{MI300X,
author = {{Advanced Micro Devices, Inc.}},
title = {AMD INSTINCT MI300X ACCELERATOR},
year = {2025},
url={https://www.amd.com/content/dam/amd/en/documents/instinct-tech-docs/data-sheets/amd-instinct-mi300x-data-sheet.pdf},
note = {retrieved 5 December, 2025}
}

@online{MI325X,
author = {{Advanced Micro Devices, Inc.}},
title = {AMD INSTINCT MI325X ACCELERATOR},
year = {2025},
url={https://www.amd.com/content/dam/amd/en/documents/instinct-tech-docs/product-briefs/instinct-mi325x-datasheet.pdf},
note = {retrieved 5 December, 2025}
}

@online{MI350X,
author = {{Advanced Micro Devices, Inc.}},
title = {AMD INSTINCT MI350X GPU},
year = {2025},
url={https://www.amd.com/content/dam/amd/en/documents/instinct-tech-docs/product-briefs/amd-instinct-mi350x-gpu-brochure.pdf},
note = {retrieved 5 December, 2025}
}

@online{MI355X,
author = {{Advanced Micro Devices, Inc.}},
title = {AMD INSTINCT MI355X GPU},
year = {2025},
url={https://www.amd.com/content/dam/amd/en/documents/instinct-tech-docs/product-briefs/amd-instinct-mi355x-gpu-brochure.pdf},
note = {retrieved 5 December, 2025}
}

@online{A100,
author = {{NVIDIA Corporation}},
title = {NVIDIA A100 Tensor Core GPU Architecture v1.0},
year = {2020},
url={https://images.nvidia.com/aem-dam/en-zz/Solutions/data-center/nvidia-ampere-architecture-whitepaper.pdf},
note = {retrieved 5 December, 2025}
}

@online{H100,
author = {{NVIDIA Corporation}},
title = {NVIDIA H100 Tensor Core GPU Architecture v1.04},
year = {2023},
url={https://resources.nvidia.com/en-us-hopper-architecture/nvidia-h100-tensor-c},
note = {retrieved 5 December, 2025}
}

@online{Blackwell,
author = {{NVIDIA Corporation}},
title = {NVIDIA Blackwell Architecture Technical Brief v2.1},
year = {2025},
url={https://resources.nvidia.com/en-us-blackwell-architecture},
note = {retrieved 5 December, 2025}
}

@online{RTX5080,
author = {{NVIDIA Corporation}},
title = {NVIDIA RTX Blackwell GPU Architecture v1.1},
year = {2025},
url={https://images.nvidia.com/aem-dam/Solutions/geforce/blackwell/nvidia-rtx-blackwell-gpu-architecture.pdf},
note = {retrieved 5 December, 2025}
}

\end{document}